\documentclass[aps,twocolumn,superscriptaddress,amsfont,amssymb,amsmath,showpacs,balancelastpage,english]{revtex4-1} 
\usepackage{babel}
\usepackage{graphicx}
\usepackage{dcolumn}
\usepackage{bm}
\usepackage{color}


\begin{document}

\title{Generalized framework for testing gravity with gravitational-wave propagation. I. Formulation}

\author{Atsushi Nishizawa}
\email{anishi@kmi.nagoya-u.ac.jp}
\affiliation{Kobayashi-Maskawa Institute for the Origin of Particles and the Universe, Nagoya University, Nagoya 464-8602, Japan}
\affiliation{Department of Physics and Astronomy, The University of 
Mississippi, University, MS 38677, USA}


\date{\today}

\begin{abstract}
The direct detection of gravitational waves (GW) from merging binary black holes and neutron stars mark the beginning of a new era in gravitational physics, and it brings forth new opportunities to test theories of gravity. To this end, it is crucial to search for anomalous deviations from general relativity in a model-independent way, irrespective of gravity theories, GW sources, and background spacetimes. In this paper, we propose a new universal framework for testing gravity with GW, based on the generalized propagation of a GW in an effective field theory that describes modification of gravity at cosmological scales. Then we perform a parameter estimation study, showing how well the future observation of GW can constrain the model parameters in the generalized models of GW propagation. 
\end{abstract}

\maketitle

\section{Introduction}

The direct detection of gravitational waves (GW) from merging binary black holes (BH) by aLIGO \cite{GW150914:detection,GW151226:detection,GW170104:detection} has demonstrated that the advanced detectors have sufficient sensitivity enough to detect GW out to the distant Universe. The fourth GW event has been observed by a detector network composed of two aLIGO and one aVIRGO for the first time \cite{GW170814:detection} and proved that three detectors can well localize the sky direction of a source. Recently a GW from binary neutron stars (NS) has been detected for the first time \cite{GW170817:detection} in coincidence with a short gamma-ray burst \cite{GW170817GRB}, followed by kilonova observations with multiple electromagnetic telescopes around the world, e.~g.~\cite{GW170817multimessenger,Coulter:2017wya,Valenti:2017ngx,Soares-Santos:2017lru,Tominaga:2017cgo}. In the coming years, the currently operating detectors will improve their sensitivities further and KAGRA will join the detector network \cite{GW-obs-plan:LRR}. It is expected that GW from the variety of compact binaries enable us to test gravity theories in strong and dynamical regimes precisely \cite{Will:2014kxa}.

To this end, it is crucial to search for anomalous deviations from general relativity (GR) in a model-independent way, because in practice it is impossible from the computational point of view to perform comprehensive GW searches in all gravity theories. One of such model-independent tests is measuring the propagation speed of a GW \cite{Will:2014kxa}. In GR, a GW propagates with the speed of light, while in an alternative theories of gravity the propagation speed could deviate from the speed of light due to the modification of gravity (see \cite{Saltas:2014dha,Bellini2014JCAP,Gleyzes2014IJMPD,Ballesteros2015JCAP} for general formulations, and for more specific cases, nonzero graviton mass \cite{Gumrukcuoglu:2011zh,DeFelice:2013nba} and extra dimensions \cite{Sefiedgar:2010we}). Also the modification of spacetime structure at a quantum level may affect the propagation of a GW \cite{AmelinoCamelia:1997gz,AmelinoCamelia:2008qg}. From the GW data of BH binaries detected by aLIGO, the constraints have been obtained on graviton mass to be $m_g < 7.7\times 10^{-23}\,{\rm{eV}}$ \cite{GW170104:detection} and on the modified dispersion relation \cite{Yunes:2016PRD,GW170104:detection}, though the latter constraint is rather weak from a theoretical point of view. Before the occurrence of the coincidence event, GW170817/GRB170817A \cite{GW170817:detection}, it had been expected that comparing arrival times between GW from a binary NS merger and high-energy photons from a short gamma-ray burst emitted almost at the same time can measure GW propagation speed at a precision of $10^{-16}$ - $10^{-15}$ \cite{Nishizawa2014PRD,Nishizawa:2016kba} and consequently tightly constrain the modification of gravity relevant to the cosmic accelerating expansion \cite{Lombriser2015arXiv}. One of the other model-independent tests is to check the existence of GW polarization modes predicted in GR and to search for additional polarizations \cite{Eardley:1973}. In GR, a GW has two polarizations, while there could be at most four additional polarizations in alternative theories of gravity. For example, in scalar-tensor theory and $f(R)$ gravity theory, additional scalar polarizations appear \cite{Maggiore:1999wm,Alves:2009eg,Alves:2010ms,Yang:2011cp}. In bimetric gravity theory and massive gravity theory, there appear at most six and five polarization modes, respectively, including scalar and vector modes \cite{Alves:2010ms,deRham:2011qq}. With multiple detectors, it is possible to detect the additional modes by separating them \cite{Nishizawa:2009bf,Hayama:2012au} or constructing a null signal \cite{Chatziioannou:2012rf}. Recently, the triple detector network of aLIGO and VIRGO has explored the existence of an additional polarization merely by showing the consistency of the detector response functions for GR polarizations when fitting to the data \cite{GW170814:detection,GW170814:polarization-pulsars}, though this is not a complete analysis based on the separation technique.

Another approach is to look for anomalous deviations from GR in the amplitude and phase of a GW waveform. Some theoretical frameworks \cite{Mishra:2010tp,Li:2011cg,Yunes:2009ke} parameterize the deviations from a GR waveform from a compact binary and the others parameterize the deviations from a GR waveform from a black-hole ringdown \cite{Gossan2012PRD,Meidam2014PRD,Glampedakis:2017dvb}. The constraints on the deviation from a GR waveform of a BH binary have been obtained in the generalized inspiral-merger-ringdown Phenom (gIMR) framework \cite{GW150914:GRtest,GW151226:details,GW170104:detection} and in the parameterized-post Einsteinian (ppE) framework \cite{Yunes:2016PRD}. These constraints aim at testing GW generation, that is, the strong regime of gravity, and are different from those aiming at GW propagation mentioned above. However, the problem of these parameterizations is that they cannot be applied to different types of GW sources such as supernovae, pulsars, stochastic background, etc. In addition, if one naively parameterizes the deviations from a GR waveform without linking to physical effects, it is difficult to interpret the physical meanings of the deviations from observations.  

To treat tests of gravity with GW more exhaustively, it is necessary to have a universally parameterized framework based on interpretable physical effects, irrespective of the models of gravity theories, GW sources, and background spacetimes. In this paper, we propose a new universal framework for testing gravity, based on the propagation equation of a GW in an effective field theory for dark energy \cite{Gleyzes2014IJMPD,Bellini2014JCAP}, which describes modification of gravity at cosmological scales, where a linear perturbation theory well holds. Then we perform a parameter estimation study, showing how well the future observation of GW can constrain the model parameters in generalized models of GW propagation. There are five advantages to focus on GW propagation. (i) The propagation equation is formulated independent of a type of GW sources (BH, NS, supernova, pulsar, stochastic background etc.) and background spacetimes (Schwarzshild, Kerr, FLRW etc.), in contrast to GW generation. The equation just describes the properties of GW propagation, independent of where the GW propagates. (ii) If one considers a different theory of gravity, the propagation properties of a GW may change. However, this deviation from GR can be easily parameterized in the propagation equation by introducing arbitrary functions that control propagation speed, amplitude damping (vacuum friction), graviton mass, and a source term (additional energy injection or escape to extra dimensions), for which physical interpretations are transparent. (iii) GW propagation allows us to test gravity in a dynamical regime at cosmological distance, at which gravity has not yet been tested precisely. The propagation of a GW itself is dynamical and the background spacetime is also dynamical due to the cosmic expansion. This regime of gravity is relevant to the origin of cosmic acceleration of the present Universe and may be related to a possible modification of GR. (iv) Even if modification on gravity is a tiny effect, propagation from a distant source can accumulate the effect and amplify a signal observed at a detector. (v) It is possible by definition to combine with the constraints from cosmological observations such as cosmic expansion, large-scale structure of the Universe, cosmic microwave background, and etc., because some of the modification functions in the propagation equation are common to those appearing in the cosmological observables, e.g.~\cite{Bellini:2015xja,Huang:2015srv,Leung:2016xli,Alonso:2016suf}.

This paper is organized as follows. In Sec.~\ref{sec2}, to develop a universal parameterized framework for testing gravity with GW propagation, we analytically solve the GW propagation equation in an effective field theory for dark energy \cite{Gleyzes2014IJMPD,Bellini2014JCAP} and obtain a WKB solution. This GW waveform is quite general because it includes arbitrary functions of time that describe modified amplitude damping, modified propagation speed, nonzero graviton mass, and a possible source term for a GW. We also show the specific expressions of these arbitrary function of gravity modifications in various alternative theories of gravity. In Sec.~\ref{sec3}, we compare our framework for generalized GW propagation with the pre-existing frameworks for testing gravity with GW, though those are relevant to GW generation. In Sec.~\ref{sec4}, we perform a parameter estimation study with a Fisher information matrix on two simple models of GW propagation whose parameters are assumed to be constant, and clarify which parameters are correlated each other and how well they are determined from realistic observational data. In Sec.~\ref{sec5}, we discuss the current constraints on the model parameters and forecast the future constraints that can be obtained by the aLIGO-like detector network at design sensitivity. Finally, Sec.~\ref{sec6} is devoted to conclusion.    

Throughout the paper, we adopt units $c=G=1$.

\section{Parameterized framework for GW propagation}
\label{sec2}

\subsection{GW propagation equation}

Following the general formulation of GW propagation in an effective field theory \cite{Saltas:2014dha}, tensor perturbations obey the equation of motion
\begin{equation}
h^{\prime\prime}_{ij}+(2+\nu){\cal H} h^{\prime}_{ij} + (c_{\rm T}^2 k^2 + a^2 \mu^2) h_{ij} = a^2 \Gamma \gamma_{ij} \;,
\label{eq:GWeq1}
\end{equation}
where the prime is a derivative with respect to conformal time, $a$ is the scale factor, ${\cal H} \equiv a^{\prime}/a$ is the Hubble parameter in conformal time, $\nu={\cal H}^{-1} (d \ln M_{*}^2/dt)$ is the Planck mass run rate, $c_{\rm T}$ is the GW propagation speed, and $\mu$ is graviton mass. The source term $\Gamma \gamma_{ij}$ arises from anisotropic stress. In the limit of $c_{\rm T}=1$ and $\nu=\mu=\Gamma=0$, the propagation equation (\ref{eq:GWeq1}) is reduced to the standard one in GR. If gravity is modified from GR, the modification functions in general depend on time and wavenumber, $\nu=\nu(\tau,k)$, $c_{\rm T}=c_{\rm T}(\tau,k)$, and $\mu=\mu(\tau,k)$. For example, if the screening mechanism works at small scales, $\nu$ can be scale-dependent. Even graviton mass can be position-dependent when an additional scalar degree of freedom is introduced, e.~g.~\cite{Zhang:2017jze}. Although the separation of $c_{\rm T}(\tau,k)$ and $\mu=\mu(\tau,k)$ is not unique, we define $\mu$ separately because we know the $k$ dependence exactly when $\mu$ is constant. 

The assumption here is that the weak equivalence principle holds for matter and therefore that all matter species external to the scalar-tensor system are coupled minimally and universally. At a linear level or large scales in the Friedmann-Lema\^{i}tre-Robertson-Walker (FLRW) background, the modification functions are simply functions of time \cite{Saltas:2014dha} (Later this is extended to allow a wave number dependence in some modified models of GW propagation). The effects of this generalized propagation of GW on the cosmic microwave background (CMB) spectrum have already been investigated numerically in \cite{Amendola:2014wma,Pettorino:2014bka,Xu:2014uba,Lin:2016gve}, though the CMB is sensitive only to modifications of gravity in the early universe, which is irrelevant to the cosmic accelerating expansion at present.

Here we focus on modifications of gravity as an explanation for the cosmic accelerating expansion and on GW observations by the second-generation detectors such as aLIGO. In other words, all the modification functions in Eq.~(\ref{eq:GWeq1}) are slowly varying functions with cosmological time scale, while GW wavelength $\sim k^{-1}$ is much smaller than the cosmological horizon scale. Thus, we can obtain a WKB solution for Eq.~(\ref{eq:GWeq1}). In the next, we derive such WKB solutions in the presence and absence of the source term $\Gamma \gamma_{ij}$.

\subsection{General case with $\Gamma=0$}
\label{sec:2-1}

Setting the solution in the form $h_{ij}=h e_{ij}=A e^{i B} e_{ij}$ with the polarization tensor $e_{ij}$ and assuming $\Gamma=0$, Eq.~(\ref{eq:GWeq1}) is reduced to the two equations:
\begin{align}
& c_{\rm T}^2 k^2 +a^2 \mu^2+ (2+\nu){\cal H} \frac{A^{\prime}}{A} -(B^{\prime})^2 + \frac{A^{\prime \prime}}{A} =0\;, \\
& (2+\nu){\cal H} + 2 \frac{A^{\prime}}{A} +\frac{B^{\prime \prime}}{B^{\prime}} =0 \;.
\end{align}
Since the case we are interested in is when modifications to gravity are slowly varying functions with a cosmological time scale, we neglect the terms $A^{\prime}/A$ and $A^{\prime \prime}/A$ in the first equation. This is justified because $c_{\rm T}^2 k^2$ and $(B^{\prime})^2$ are quantities in GW phase and change with the time scale of GW period, while $A^{\prime}/A$ and $A^{\prime \prime}/A$ are of the order of ${\cal H}^2$, which is much smaller than $k^2$. In addition, we know from GW observations \cite{GW150914:GRtest} that graviton mass is smaller than $1.2 \times 10^{-22}\,{\rm eV}$. Then the condition $a^2 \mu^2/k^2 \ll c_{\rm T}^2 \sim 1$ is always satisfied for GW detectors in the late-time cosmology and guarantees a wavy solution. From the first equation, the phase part is
\begin{equation}
B = \pm k \int^{\tau} \tilde{c}_{\rm T} \,d\tau^{\prime} \;. 
\end{equation}
where $\tilde{c}_{\rm T}^2 \equiv c_{\rm T}^2+ a^2 \mu^2/k^2$ is an effective GW speed and the $\tau$ integral runs from GW emission time at a source to detection time at the Earth. Substituting this for the second equation, we have the WKB solution
\begin{align}
h &\propto \frac{q}{\sqrt{\tilde{c}_{\rm T}}} \exp \left[ \pm i k \int^{\tau} \tilde{c}_{\rm T} \, d\tau^{\prime} \right] \;, 
\label{eq10} \\
q &\equiv \exp \left[ -\int^{\tau} \left( 1+ \frac{\nu}{2}  \right) {\cal H} \, d\tau^{\prime} \right] \;.
\end{align}

To separate the correction due to gravity modification, we define $c_{\rm T} \equiv 1-\delta_g$. For simplification, we replace $\tilde{c}_{\rm T}$ with $c_{\rm T}$ in amplitude of Eq.~(\ref{eq10}) because $\delta_g$ and $a \mu/k$ can be tightly constrained from the phase correction. Indeed $a \mu/k$ has already been limited to be much smaller than unity from LIGO observations \cite{GW150914:GRtest}. Then the WKB solution is
\begin{align}
h &\propto \exp \left[- \frac{1}{2} \int^{\tau} \nu  {\cal H} d\tau^{\prime} \right] \exp \left[ \mp i k \int^{\tau} \left( \delta_g -\frac{a^2 \mu^2}{2k^2} \right) d\tau^{\prime} \right] \nonumber \\
&\times \exp \left[ -\int^{\tau} {\cal H} d\tau^{\prime} \right]  \exp \left[ \pm i k \int^{\tau} d\tau^{\prime} \right] \;. \label{eq2} 
\end{align}
Since the last two exponential factors appear in GR, the WKB solution can be written a more transparent way by factorizing out a GR waveform, assuming GW generation is the same as in GR. The sign of phase is defined by the GR waveform phase in Eq.~(\ref{eq:PNphase}) and we must choose the upper sign in Eq.~(\ref{eq2}). Finally, the waveform is expressed as
\begin{align}
h &= {\cal{C}}_{\rm MG} h_{\rm GR} \;, 
\label{eq3} \\
{\cal{C}}_{\rm MG} &\equiv e^{- {\cal D}} e^{- i k \Delta T} \;,
\end{align}
with
\begin{align}
{\cal D} &\equiv \frac{1}{2} \int^{\tau} \nu  {\cal H} d\tau^{\prime} \nonumber \\
&= \frac{1}{2} \int_0^z \frac{\nu}{1+z^{\prime}} dz^{\prime} \;, 
\end{align}
\begin{align}
\Delta T &\equiv \int^{\tau} \left( \delta_g -\frac{a^2 \mu^2}{2k^2} \right) d\tau^{\prime} \nonumber \\
&= \int_0^z \frac{1}{\cal H} \left( \frac{\delta_g}{1+z^{\prime}} -\frac{\mu^2}{2k^2 (1+z^{\prime})^3} \right) dz^{\prime} \;.
\label{eq:time-delay}
\end{align}
where ${\cal D}$ is the damping factor, and $\Delta T$ is the time delay due to the effective GW speed different from speed of light. We call this solution the generalized GW propagation (gGP) framework to test gravity. It is quite general and can be applied to many theories of modified gravity such as Horndeski theory, including $f(R)$ gravity as a special case, and Einstein-aether theory. Following the classification in \cite{Saltas:2014dha}, concrete expressions for each modification function are listed in Table~\ref{tab1}.

\begin{table*}[t]
\begin{center}
\begin{tabular}{cccccc}
\hline \hline
gravity theory & $\nu$ & $c_{\rm T}^2-1$ & $\mu$ & $\Gamma$ & Refs. \\
\hline 
&&&&& \\
general relativity & 0 & 0 & 0 & 0 & --- \\
&&&&& \\
extra-dim. theory & $(D-4) \left( 1+ \frac{1+z}{{\cal{H}}d_{\rm L}} \right)$ & 0 & 0 & 0 & Sec.~\ref{sec2-4} \\
&&&&& \\
Horndeski theory & $\alpha_M$ & $\alpha_T$ & 0 & 0 & \cite{Saltas:2014dha} \\
&&&&& \\
f(R) gravity & $F^{\prime}/{\cal H} F$ & 0 & 0 & 0 & \cite{Hwang:1996xh} \\
&&&&& \\
Einstein-aether theory & 0 & $c_{\sigma}/(1+c_{\sigma})$ & 0 & 0 & \cite{Saltas:2014dha} \\
&&&&& \\
modified dispersion relation & 0 & $(n_{\rm mdr}-1) \mathbb{A} E^{n_{\rm mdr}-2}$ & when $n_{\rm mdr}=0$ & 0 & \cite{Yunes:2016PRD} \\
&&&&& \\
bimetric massive gravity theory & 0 & 0 & $m^2 f_1$ & $m^2 f_1$ & \cite{Saltas:2014dha,Max:2017flc} \\
&&&&& \\
\hline \hline
\end{tabular}
\end{center}
\caption{Modification functions, $\nu$, $c_{\rm T}$, $\mu$, and $\Gamma$ in specific modified gravity theories. In the phenomenology of a modified dispersion relation, a special case with $n_{\rm mdr}=0$ gives nonzero graviton mass.}
\label{tab1}
\end{table*}

\begin{figure*}[t]
\begin{center}
\includegraphics[width=5.5cm]{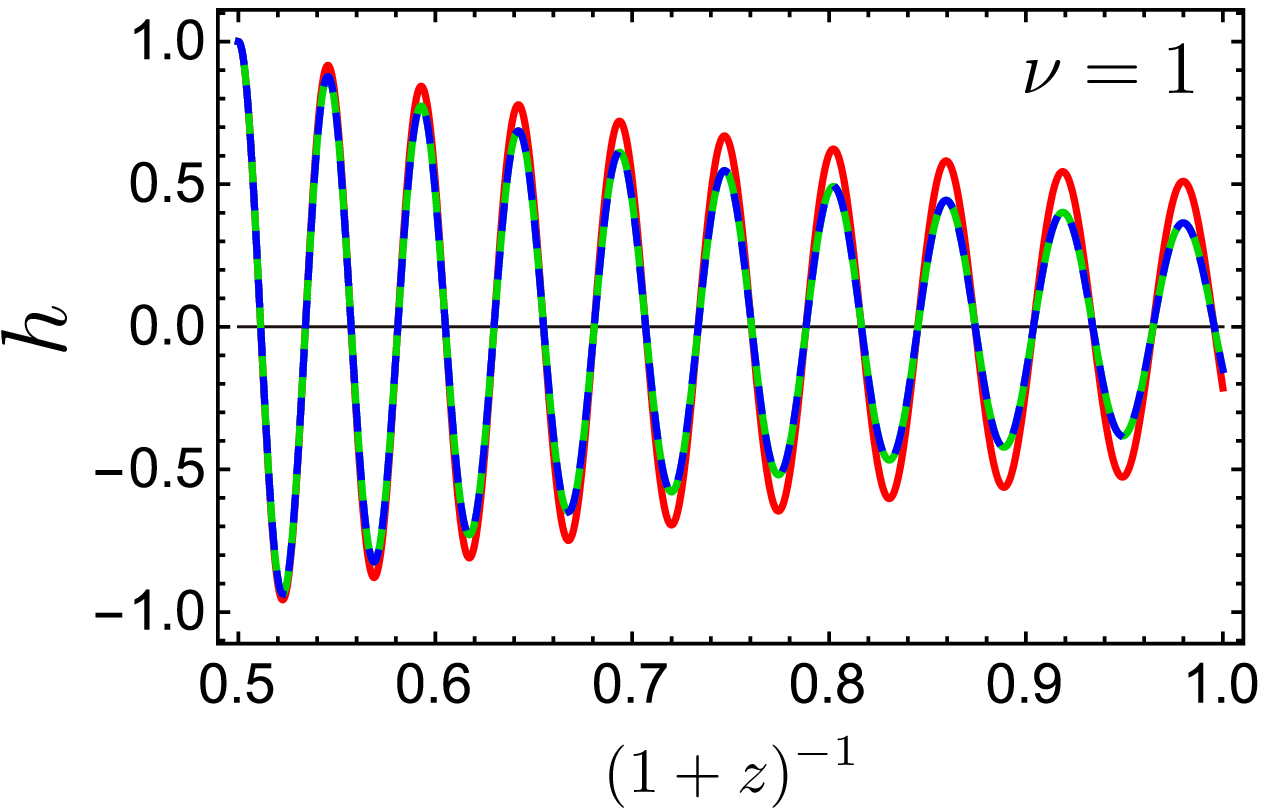}
\includegraphics[width=5.5cm]{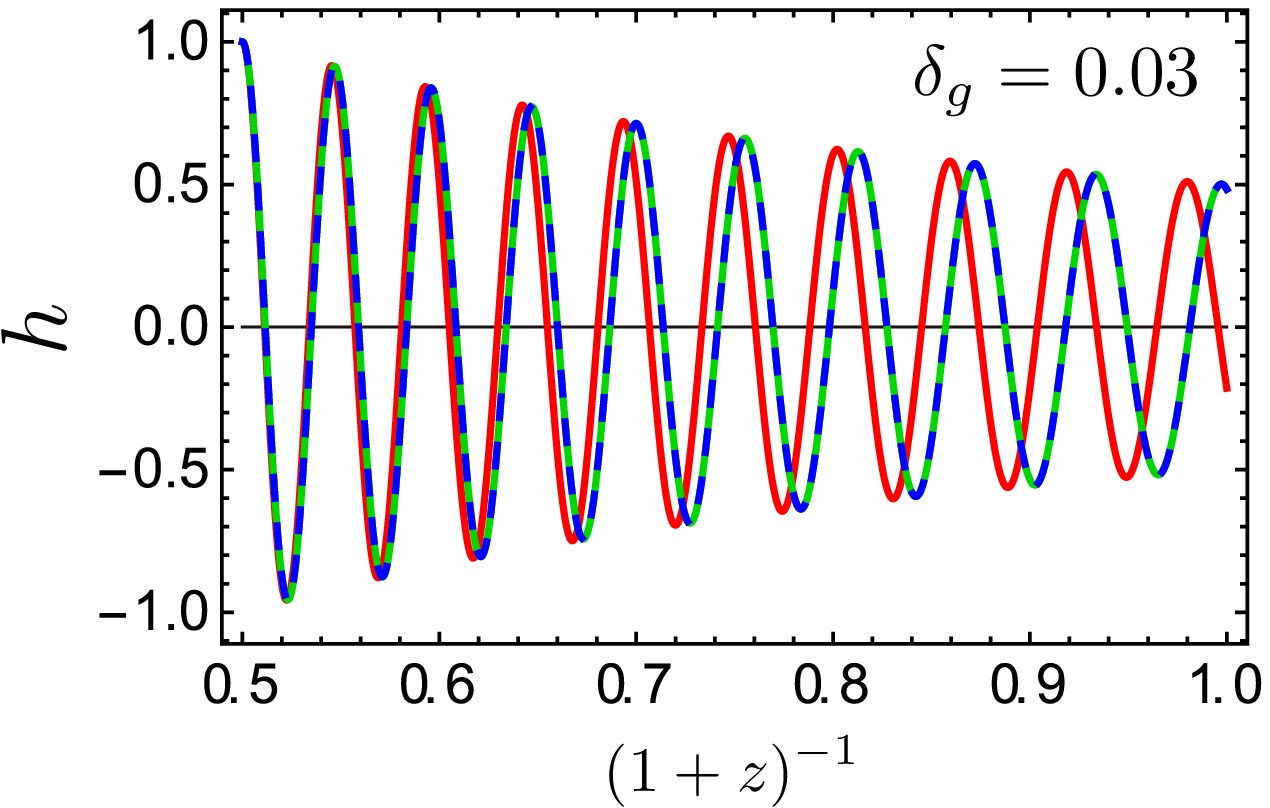}
\includegraphics[width=5.5cm]{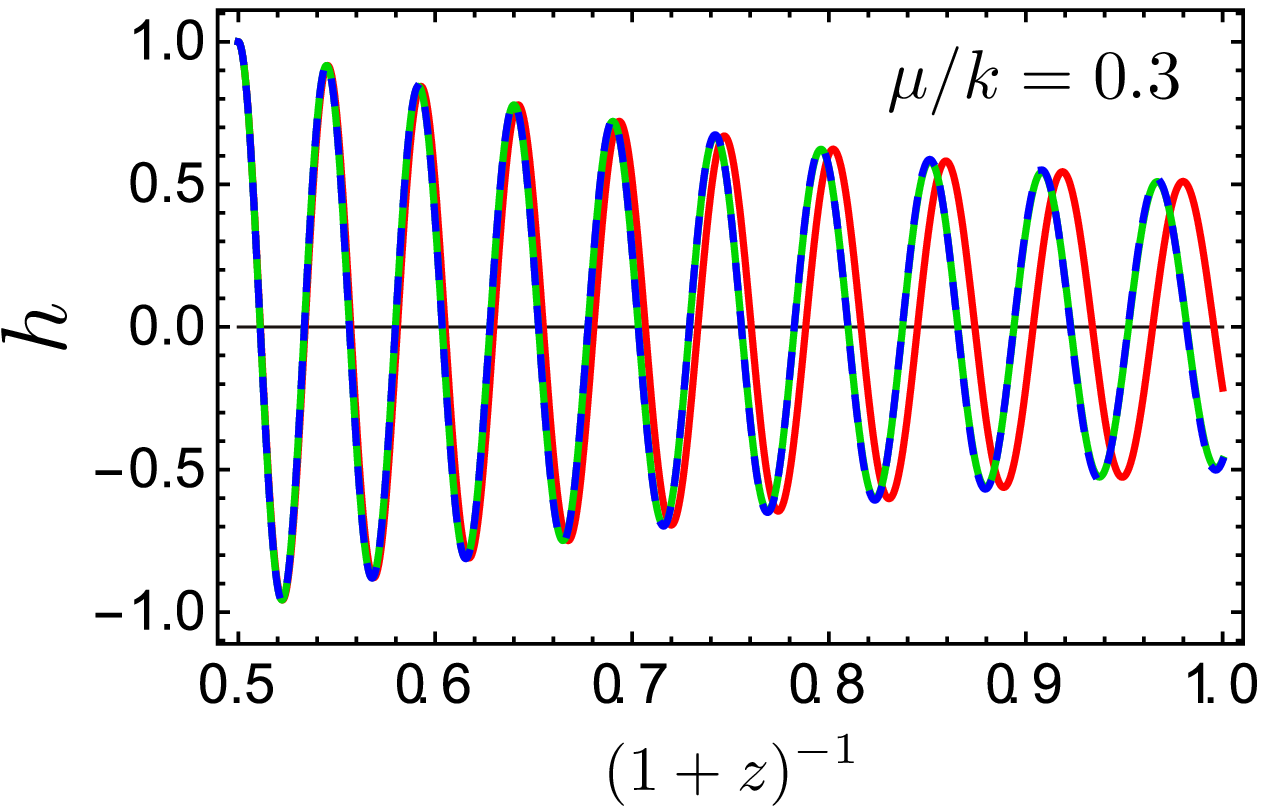}
\caption{Modified GW waveforms in the Einstein-de Sitter universe ($\Omega_{\rm m}=1$) with $\nu=1$ (left), $\delta_g=0.03$ (middle), $\mu/k=0.3$ (right). In each panel, only one parameter is changed. The curves are a GR waveform (red, solid), a numerical modified waveform (green, solid), and a WKB modified waveform (blue, dashed). The wave number is fixed to $\tilde{k} \equiv k \tau_0=200$ and the initial condition is set so that $h=1$ and $h^{\prime}=0$ at $z=1$ just for illustration.}
\label{fig2}
\end{center}
\end{figure*}

Particularly, when all arbitrary functions $\nu, c_{\rm T}, \mu$ are assumed to be constant and $\Gamma=0$, the WKB solution in Eq.~(\ref{eq3}) is significantly simplified as
\begin{align}
h &= (1+z)^{-\nu/2} e^{- i k \Delta T}  h_{\rm GR} 
\label{eq9} \;, \\
\Delta T &= \frac{\delta_g d_{\rm L}}{1+z}  - \frac{\mu^2}{2k^2} \int_{0}^{z} \frac{dz^{\prime}}{(1+z^{\prime})^3 {\cal H}} \;,
\end{align}
where
\begin{equation}
d_{\rm L}(z) = (1+z) \int_{0}^{z} \frac{dz^{\prime}}{(1+z^{\prime}){\cal H}} \;.
\label{eq:dL}
\end{equation}
Examples of a modified GW waveform are shown in Fig.~\ref{fig2}.

\subsection{General case with $\Gamma \neq 0$}
\label{sec2-2}

If the propagation equation in Eq.~(\ref{eq:GWeq1}) is inhomogeneous ($\Gamma \neq 0$), a solution becomes much more complicated, but can be formally obtained. Denoting homogeneous solutions by
\begin{align}
u_1 (\tau) &\equiv \frac{q(\tau)}{\sqrt{\tilde{c}_{\rm T} (\tau)}} \cos \left[ k \int^{\tau} \tilde{c}_{\rm T} d\tau^{\prime} \right] \;, \\
u_2 (\tau) &\equiv \frac{q(\tau)}{\sqrt{\tilde{c}_{\rm T} (\tau)}} \sin \left[ k \int^{\tau} \tilde{c}_{\rm T} d\tau^{\prime} \right] \;,
\end{align}
and using the following relations
\begin{align}
&u_1(\tau^{\prime}) u_2(\tau) - u_1(\tau) u_2(\tau^{\prime}) \nonumber \\
& \quad \quad = \frac{q(\tau) q(\tau^{\prime})}{\sqrt{\tilde{c}_{\rm T} (\tau)\tilde{c}_{\rm T}(\tau^{\prime})}} \sin \left[ k \int_{\tau^{\prime}}^{\tau} \tilde{c}_{\rm T} d\tau^{\prime\prime}  \right] \;, \\
&u_1(\tau^{\prime}) u_2^{\prime}(\tau^{\prime}) - u_1^{\prime}(\tau^{\prime}) u_2(\tau^{\prime}) = k q^2(\tau^{\prime}) \;,
\end{align}
an inhomogeneous solution is
\begin{widetext}
\begin{align}
h_{ij}(\tau) &=\left\{ C_1 u_1(\tau) + C_2 u_2 (\tau) \right\} e_{ij} + \int^{\tau} a^2(\tau^{\prime}) \Gamma(\tau^{\prime}) \gamma_{ij} (\tau^{\prime}) \frac{u_1(\tau^{\prime}) u_2(\tau) - u_1(\tau) u_2(\tau^{\prime})}{ u_1(\tau^{\prime}) u_2^{\prime}(\tau^{\prime}) - u_1^{\prime}(\tau^{\prime}) u_2(\tau^{\prime})} d\tau^{\prime} \nonumber \\
&=\left\{ C_1 u_1(\tau) + C_2 u_2 (\tau) \right\} e_{ij} + \frac{q(\tau)}{k \sqrt{\tilde{c}_{\rm T} (\tau)}} \int^{\tau} \frac{a^2(\tau^{\prime}) \Gamma(\tau^{\prime}) \gamma_{ij} (\tau^{\prime})}{q (\tau^{\prime}) \sqrt{\tilde{c}_{\rm T} (\tau^{\prime})}} \sin \left[ k \int_{\tau^{\prime}}^{\tau} \tilde{c}_{\rm T} d\tau^{\prime\prime}  \right] d\tau^{\prime} \;.
\label{eq:inhom-sol}
\end{align}
\end{widetext}
where $C_1$ and $C_2$ are arbitrary coefficients.

The existence of nonzero $\Gamma$ modifies GW amplitude as $\Gamma$ behaves as a source term for a GW. The simplest case is GW propagation in the standard cosmology with anisotropic stress $\pi_{ij}$, but without modifying gravity \cite{Weinberg:2003ur}. Setting the model parameters to $c_{\rm T}=1$, $\nu=0$, $\mu=0$, and $\Gamma \gamma_{ij} = 16 \pi \pi_{ij}$ and replacing the integrals with
\begin{equation}
\int_{\tau^{\prime}}^{\tau} d\tau^{\prime\prime} = \int_{z^{\prime}}^z \frac{dz^{\prime \prime}}{H(z^{\prime \prime})} \;, \nonumber
\end{equation} 
the inhomogeneous solution is expressed as
\begin{align}
h_{ij} &= \left\{ C_1 u_1(\tau) + C_2 u_2 (\tau) \right\} e_{ij} \nonumber \\
&+ \frac{1}{1+z} \int_0^{z} \frac{\Gamma (z^{\prime}) \gamma_{ij} (z^{\prime})}{(1+z^{\prime}) H(z^{\prime}) k} \sin \left[ k \int_{z^{\prime}}^z \frac{dz^{\prime \prime}}{H(z^{\prime \prime})} \right] dz^{\prime} \;,
\end{align}
where $H(z) = (1+z) {\cal H}(z)$. Since the phase of the sine function is $k(\tau-\tau^{\prime}) \gg 1$ when one considers GW frequency relevant to GW detectors, the integrand is rapidly oscillating, changing its sign. However, the magnitude of correction to GW amplitude is of the order of $\Gamma \gamma_{ij}/H k$ and is roughly proportional to propagation distance. 

In the bimetric gravity theory, the model parameters are $c_{\rm T}=1$, $\nu=0$, $\mu=m^2 f_1$, and $\Gamma \gamma_{ij} = m^2 f_1 \gamma_{ij}$. In addition to the source term, graviton mass has nonzero value with the same dependence as $\Gamma$. Although the correction term is more complicated, the modification of GW amplitude is similar to that in the simple case with anisotropic stress.

\subsection{Modified dispersion relation}
\label{sec2-3}

In phenomenological models of quantum gravity, quantum fluctuations of spacetime can modify the dispersion relation of a massless particle at a low energy limit $E \ll E_{\rm QG}$ \cite{AmelinoCamelia:1997gz}
\begin{equation}
E^2 = p^2 \left[ 1+ \xi \left( \frac{E}{E_{\rm QG}} \right)^{n_{\rm QG}-2} \right]  \;.
\label{eq:MDR}
\end{equation}
In addition, similar modification of dispersion relation can be introduced by Lorentz invariance violation, nonzero graviton mass, and extra dimensions \cite{Kostelecky:2016PLB,Will:1997bb,Sefiedgar:2010PLB}. Consequently, GW speed depends on graviton energy or GW frequency. Here we extend the gGP framework by allowing a wave number dependence for GW propagation speed. The GW propagation speed (phase velocity) is derived from Eq.~(\ref{eq:MDR}).
\begin{equation}
c_{\rm T}(E) \equiv \frac{E}{p} \approx 1+ \frac{\xi}{2} \left( \frac{E}{E_{\rm mdr}} \right)^{n_{\rm mdr}-2} \;.
\label{eq5} 
\end{equation}
Here we denote $n_{\rm QG}$ and $E_{\rm QG}$ by $n_{\rm mdr}$ and $E_{\rm mdr}$, taking into account modifications of the dispersion relation other than quantum gravity effect. While graviton speed (group velocity) is
\begin{equation}
v_{\rm g}(E) \equiv \frac{dE}{dp} \approx 1+ \frac{\xi}{2} (n_{\rm mdr}-1) \left( \frac{E}{E_{\rm mdr}} \right)^{n_{\rm mdr}-2} \;.
\end{equation}
Defining $\mathbb{A} \equiv \xi E_{\rm mdr}^{2-n_{\rm mdr}}$, modifications on group velocity are summarized in the form \cite{Mirshekari2012PRD,Yunes:2016PRD},
\begin{equation}
v_{\rm g}(E) =1+\frac{(n_{\rm mdr}-1)}{2} \mathbb{A} E^{n_{\rm mdr}-2} \;.
\end{equation}
Note that $\mathbb{A}^{1/(2-n_{\rm mdr})}$ is roughly equivalent to characteristic energy scale of a theory $E_{\rm mdr}$, at which a quantum gravity effect is switched on or a graviton starts to be sensitive to extra dimensions. The amplitude $\mathbb{A}$ and the power index $n_{\rm mdr}$ in specific gravity theories are listed in Table~\ref{tab2}.

In terms of our formulation, other properties of a GW are not modified by the modification of dispersion relation, that is, $\nu=0$, $\mu=0$ (except for $n_{\rm mdr}=0$), and $\Gamma=0$. Using $\delta_g$, Eq.~(\ref{eq5}) is expressed as 
\begin{align}
\delta_g &\approx - \frac{1}{2} \mathbb{A} E^{n_{\rm mdr}-2} \nonumber \\
&= - \frac{(2\pi)^{n_{\rm mdr}-2}}{2} \mathbb{A} f^{n_{\rm mdr}-2} \;,
\end{align}
and is related to $v_{\rm g}$ by
\begin{equation}
\delta_g = \frac{1-v_{\rm g}}{n_{\rm mdr}-1} \quad \quad (n_{\rm mdr} \neq 1) \;.
\label{eq7}
\end{equation}
Note that this relation is valid only when the form of the modified dispersion relation in Eq.~(\ref{eq:MDR}) is assumed.

\begin{table}[h]
\begin{center}
\begin{tabular}{cccc}
\hline \hline
gravity theory & $\mathbb{A}$ & $n_{\rm mdr}$ & Refs. \\
\hline  
&&& \\
general relativity & 0 & --- & --- \\
&&& \\
massive graviton & $(\mu c^2)^2$ & 0 & \cite{Will:1997bb} \\
&&& \\
doubly special relativity & $\eta_{\rm dsrt}$ & 3 & \cite{AmelinoCamelia:2000PLB} \\
&&& \\
extra-dimensional theories & $-\alpha_{\rm edt}$ & 4 & \cite{Sefiedgar:2010PLB} \\
&&& \\
Horava-Lifshitz gravity & $k_{\rm hl}^4 \mu_{\rm hl}^2/16$ & 4 & \cite{Mukohyama:2009PLB,Vacaru:2010GRG}\\
&&& \\
gravitational SME (even $d\geq 4$) & $-2 \mathring{k}_{(V)}^{(d)}$ & d-2 & \cite{Kostelecky:2016PLB} \\
&&& \\
gravitational SME (odd $d\geq 5$) & $\pm 2\mathring{k}_{(V)}^{(d)}$ & d-2 & \cite{Kostelecky:2016PLB} \\
&&& \\
\hline \hline 
\end{tabular}
\end{center}
\caption{Modified dispersion relations in specific modified gravity theories. SME stands for standard model extensions.}
\label{tab2}
\end{table}

We add a caveat on violation of the weak equivalence principle and GW propagation speed. If the weak equivalence principle is violated, gravitons with different energy or frequency trace different null geodesics, responding differently to gravitational potential along the line of sight. Then these gravitons arrive at the Earth at different times even if they are emitted simultaneously at a source and propagate with the speed of light ($c_{\rm T}=1$). In the gGP framework here, we assume that the weak equivalence principle holds for matter and the propagation speed is exactly $c_{\rm T}=1$ when other modifications on gravity is absent. The constraints on the violation of the weak equivalence principle have been obtained from GW observations in \cite{Kahya:2016prx,DeLaurentis:2016jfs}.

\subsection{Extra-dimensional theory}
\label{sec2-4}

In a universal extra-dimensional theory, a GW damps with $h \propto d_{\rm L}^{-(D-2)/2}$ due to leakage to extra dimensions \cite{Cardoso:2002PRD}. Namely, in a higher dimensional spacetime with $D>4$, a GW damps faster than in $D=4$ spacetime. If there exists a crossover distance scale $R_{\rm c}$ beyond which spacetime dimension behaves differently, GW amplitude scales with
\begin{equation}
h \propto \left[ d_{\rm L} \left\{ 1+ \left( \frac{d_{\rm L}}{R_{\rm c}} \right)^{n_{\rm c}/2} \right\}^{(D-4)/n_{\rm c}}  \right]^{-1} \;,
\end{equation}   
where the power index $n_{\rm c}$ represents transition steepness, as proposed in \cite{Deffayet:2007ApJL}. Then the correspondence to our formulation when $d_{\rm L} \gg R_{\rm c}$ is
\begin{equation}
e^{- {\cal D}} = \exp \left[ -\frac{1}{2} \int_{z_{\rm c}}^z \frac{\nu}{1+z^{\prime}} dz^{\prime} \right] = \left( \frac{d_{\rm L}}{R_{\rm c}} \right)^{-(D-4)/2} \;,
\end{equation}
where $z_{\rm c}$ is the redshift corresponding to $R_{\rm c}$. Again this is the same damping as the universal extra-dimensional theory, $h \propto d_{\rm L}^{-(D-2)/2}$.

Here we connect the effect of extra-dimensions to $\nu$. Of course, in general, the contribution to $\nu$ is not only from extra-dimensions but from modification of gravity strength itself. In order to distinguish them, we additionally define $\nu_{\rm ext}$, which is solely due to the extra-dimensional effect. After some algebra with Eq.~(\ref{eq:dL}), we have
\begin{equation}
\nu_{\rm ext} = (D-4) \left( 1+ \frac{1+z}{ {\cal H} d_{\rm L}} \right) \;.
\end{equation}
When $D=4$, there is no extra amplitude damping and the standard damping, $\nu=0$, is recovered. At a large distance, it approaches $\nu=D-4$.

\subsection{Extra polarizations}

In the generalized propagation framework above, we concentrated only on the tensor mode of a GW. However, if a GW is produced via parity-violating process and has chirality, the GW may have the properties different from GR for plus and cross polarizations, e.g.~polarization-dependent propagation speed or anomalous amplitude ratio. As a result, the GW has linear or circular polarizations. While in some modified theories of gravity, there may exist additional polarizations corresponding to new degrees of freedom in the theories, e.g.~scalar and vector modes \cite{Eardley:1973,Will:book}. In all these cases, if each polarization mode decouples, one can write down propagation equations similar to Eq.~(\ref{eq:GWeq1}) for each polarization mode and introduce other families of modified gravity parameters in each GW waveform.

\section{Relations to other parameterized frameworks}
\label{sec3}

There has been no parameterized framework aiming at GW propagation. However, several parameterized frameworks for compact binary coalescence in a strong gravity regime have been proposed and enable us to compare GW propagation effects with GW generation effects. Of course, observational data include the effects of both GW generation and propagation. However, comparing both effects gives some insights into how generation and propagation effects of GW are distinguished in observational data. In this section, we compare our gGP framework with two other frameworks for GW generation: the ppE model and the gIMR model. Then we derive relations between model parameters in different frameworks and show that the propagation effects can be distinguished from the generation effects.

\subsection{parameterized-post Einsteinian framework}
In the ppE framework \cite{Yunes:2009ke}, a GW waveform is parameterized by
\begin{equation}
h(f) = \left(1+ \sum_j \alpha_j u^j \right) e^{i \sum_k \beta_k u^k} h_{\rm GR}(f) \;, \\
\end{equation} 
where $u \equiv (\pi {\cal M} f)^{1/3}$. GR is recovered at the limit of $\alpha_j \rightarrow 0$ and $\beta_k \rightarrow 0$. Compared with the gGP framework in Eq.~(\ref{eq3}), the following relations hold.
\begin{align}
\sum_j \alpha_j u^j &= - \frac{1}{2} \int_0^z \frac{\nu}{1+z^{\prime}} dz^{\prime} \;, \\
\sum_k \beta_k u^k &= - k \int_0^z \left( \frac{\delta_g}{1+z^{\prime}} - \frac{\mu^2}{2k^2 (1+z^{\prime})^3} \right) \frac{dz^{\prime}}{\cal{H}} \;.
\end{align}

Since $u$ gives a specific frequency dependence, the above relations are simplified only if $\delta_g$ and $\nu$ do not depend on $k$ ($\mu$ is independent of $k$ by definition),
\begin{align}
\alpha_0 &= -\frac{1}{2} \int_0^z \frac{\nu}{1+z^{\prime}} dz^{\prime} \;, \\
\beta_3 &= - \frac{2}{\cal{M}} \int_0^z \frac{\delta_g}{(1+z^{\prime}){\cal{H}}} dz^{\prime} \;, \\
\beta_{-3} &= \frac{\cal{M}}{2} \int_0^z \frac{\mu^2}{(1+z^{\prime})^3 {\cal{H}}}dz^{\prime} \;. 
\end{align}
In terms of parameterized-post Newtonian formalism, $\nu$ correction is Newtonian-order in amplitude, $\delta_g$ correction is 4 post-Newtonian (PN) order in phase, and $\mu$ correction is 1 PN order in phase. This does not necessarily mean that higher PN effects are small, because these effects are accumulated during propagation and are amplified, proportional to propagation distance. In other words, these higher PN terms for propagation in principle could exceed the standard PN terms at the lower orders for wave generation. We note that the PN order for wave propagation is nothing to do with the PN expansion, but merely refers to frequency dependence. 

If $\nu$ and $\delta_g$ have a specific dependence on wavenumber or frequency, the corresponding PN terms change. Extending $\nu$ and $\delta_g$ in power of $k$ with a characteristic scale $k_0$, 
\begin{align}
\nu (k) &= \nu^{(0)} +  \nu^{(1)} \left( \frac{k}{k_0} \right) +  \nu^{(2)} \left( \frac{k}{k_0} \right)^2 + \cdots \;, \\ 
\delta_g (k) &= \delta_g^{(0)} + \delta_g^{(1)} \left( \frac{k}{k_0} \right) + \delta_g^{(2)} \left( \frac{k}{k_0} \right)^2 + \cdots \;,
\end{align}
we have the relations
\begin{align}
\alpha_{3j} &= -\frac{1}{2} \left( \frac{2}{{\cal M}k_0} \right)^j \int_0^z \frac{\nu^{(j)}}{1+z^{\prime}} dz^{\prime} \;, \\
\beta_{3(j+1)} &= - \frac{2}{\cal{M}} \left( \frac{2}{{\cal M}k_0} \right)^j \int_0^z \frac{\delta_g^{(j)}}{(1+z^{\prime}){\cal{H}}} dz^{\prime} \;, 
\end{align}
where $j=0,1,2,\cdots$.
Note that negative powers of $k$ is not allowed to guarantee the well-behaved low-energy limit. The coefficients $\alpha_{3j}$ and $\beta_{3(j+1)}$ correspond to $1.5j$ PN order in amplitude and $(4+1.5j)$ PN order in phase, respectively.

\subsection{generalized IMR Phenom framework}

The gIMR framework \cite{Li:2011cg} is a subclass of the ppE framework, which is used recently by LIGO scientific collaboration to test gravity in a strong field regime \cite{GW150914:GRtest,GW151226:details,GW170104:detection}. This model includes deviations from GR only in GW phase and is parameterized as
\begin{equation}
h(f) = e^{i \delta \Phi_{\rm gIMR}} h_{\rm GR}(f) \;,
\end{equation} 
where
\begin{equation}
\delta \Phi_{\rm gIMR} = \frac{3}{128\eta} \sum_{i=0}^{7} \phi_i \delta \chi_i (\pi M f)^{(i-5)/3} \;, 
\end{equation}
and $M$ is total mass, $\eta=m_1 m_2/(m_1+m_2)^2$ is symmetric mass ratio, and $\phi_i$ is the $i$-th order post-Newtonian (PN) phase in GR \cite{Khan:2016PRD}. The relation to the gGP framework is
\begin{align}
& \frac{3}{128\eta} \sum_{i=0}^{7} \phi_i \delta \chi_i (\pi M f)^{(i-5)/3} \nonumber \\
&=- k \int_0^z \left( \frac{\delta_g}{1+z^{\prime}} - \frac{\mu^2}{2k^2 (1+z^{\prime})^3} \right) \frac{dz^{\prime}}{\cal{H}} \;.
\end{align}
Here $\nu$ is irrelevant to the gIMR model because no amplitude correction is considered. 

If $\delta_g$ does not depend on $k$ ($\mu$ is independent of $k$ by definition), there are simple relations
\begin{align}
\delta \chi_8 &= - \frac{256\eta}{3 M \phi_8} \int_0^z \frac{\delta_g}{1+z^{\prime}} dz^{\prime} \;, \\
\delta \chi_2  &=  \frac{32M \eta}{3\phi_2} \int_0^z \frac{\mu^2}{(1+z^{\prime})^3 {\cal{H}}}dz^{\prime} \;.
\end{align}
However, the 4 PN phase in GR, $\phi_8$, is not completely known yet. Therefore, we cannot connect $\delta_g$ to the gIMR model exactly. In addition, $\nu$ correction is out of this gIMR framework because amplitude modification is not considered by definition.

\subsection{Generation effect vs propagation effect}

In the above, we naively connected the gGP framework to other frameworks and derived their correspondences. However, from the observational point of view, data from detectors include both generation and propagation effects and we need to distinguish them. There are three reasons why we assume that a generation effect is ignored in the gGP framework. First, a degeneracy between generation and propagation effects is problematic only when they are at the same PN order (with the same frequency dependence). Although various theories that could alter GW generation are listed in \cite{Yunes:2016PRD,Chamberlain:2017fjl}, all effects in gravity modification come in at the order lower than 2 PN in phase. On the other hand, propagation effects come in at higher PN order than 4 PN except for graviton mass at 1 PN. Second, as discussed in \cite{Yunes:2016PRD}, a generation effect is in general much smaller because a propagation effect is accumulated, proportional to propagation distance. Third, most importantly, a propagation effect increases proportional to source distance and can in principle be distinguished by analyzing multiple sources. The last point has been demonstrated in \cite{Nishizawa:2016kba}, distinguishing the modification effect of GW propagation speed from intrinsic emission time delay at a source. The above reason also indicates that tighter constraints can be obtained once generation and propagation tests of gravity are combined.

\section{Parameter estimation from GW observations}
\label{sec4}

In this section, we investigate a simple model in Eq.~(\ref{eq9}), in which arbitrary functions $\nu, c_{\rm T}, \mu$ are assumed to be constant and $\Gamma=0$. Using this waveform, we demonstrate with a Fisher information matrix how precisely we can measure the model parameters from realistic observations of GW. 

\subsection{GW waveform}
\label{sec4a}

For the GR waveform, $h_{\rm GR}$, we will use the phenomenological waveform (PhenomD) \cite{Khan:2016PRD}, which is an up-to-date version of inspiral-merger-ringdown (IMR) waveform for aligned-spinning (nonprecessing) BH-BH binaries with mass ratio up to 1:18. While for BH-NS and NS-NS binaries, we will use the inspiral waveform up to 3.5 PN order in phase, which is an early inspiral part of the PhenomD waveform. This is because tidal deformation and disruption of a NS prevent us from analytically modeling the merger phase for a NS binary and from observing a clean ringdown signal after the merger. 

The PhenomD waveform is composed of three parts: inspiral, intermediate, and merger-ringdown phases. The explicit expressions are given in Appendix \ref{app:IMR}, but the overall structure is given as follows:  
\begin{equation}
h_{\rm GR} = {\cal G}_I A_{\rm IMR}\, e^{i \phi_{\rm IMR}} \;, 
\label{eq:PhenomD-full} \\
\end{equation}
\begin{equation}
A_{\rm IMR} =  \left\{
\begin{array}{ll|} 
\displaystyle
A_{\rm ins}  \quad \quad f \leq f_{a1}
\\ \\
\displaystyle 
A_{\rm int} \quad \quad f_{a1} < f \leq f_{a2}
\\ \\
\displaystyle 
A_{\rm MR} \quad \quad f_{a2} < f
\end{array}
\right. \;, 
\end{equation}
\begin{equation}
\phi_{\rm IMR} =  \left\{
\begin{array}{ll|} 
\displaystyle
\phi_{\rm ins,E} + \phi_{\rm ins,L}  \quad f \leq f_{p1}
\\ \\
\displaystyle 
\phi_{\rm int} \quad \quad f_{p1} < f \leq f_{p2}
\\ \\
\displaystyle 
\phi_{\rm MR} \quad \quad f_{p2} < f
\end{array}
\right. \;. 
\end{equation}
Here ${\cal G}_I$ is the geometrical factor for $I$-th detector defined by 
\begin{align}
{\cal G}_I &\equiv \left\{ \frac{ 1+ \cos^2 \iota}{2} F_{+,I}(\theta_{\rm S}, \phi_{\rm S}, \psi)  + i \cos \iota\, F_{\times,I}(\theta_{\rm S}, \phi_{\rm S}, \psi) \right\} \nonumber \\
 & \quad \times e^{- i \phi_{\rm D, I}(\theta_{\rm S}, \phi_{\rm S}) } \;,
\end{align}
where $\phi_{\rm D,I}$ is the Doppler phase for $I$-th detector, and $F_{+,I}$ and $F_{\times,I}$ are $I$-th detector's response functions to each polarization mode, respectively, e.~g.~\cite{Cutler:1994ys}. Note that the transition frequencies do not coincide exactly for amplitude and phase. The waveform of a simple model in Eq.~(\ref{eq9}) has in total 14 parameters: the redshifted chirp mass ${\cal{M}}$, the symmetric mass ratio $\eta$, time and phase at coalescence, $t_{\rm c}$ and $\phi_c$, redshift $z$, symmetric and asymmetric spins, $\chi_s$ and $\chi_a$, the angle of orbital angular momentum measured from the line of sight $\iota$, sky direction angles of a source, $\theta_{\rm S}$ and $\phi_{\rm S}$, polarization angle $\psi$, and gravitational modification parameters, $\delta_g$, $\nu$, and $\mu$. In a simple model, modified gravity parameters are $\delta_g$, $\nu$, and $\mu$ and are assumed to be constant. In addition, for simplicity we will assume a flat Lambda-Cold-Dark-Matter ($\Lambda$CDM) model and fix cosmological parameters to those determined by Planck satellite \cite{Planck2015cosmology}. This is justified because we are interested in the models that explain the accelerating expansion of the universe at low redshifts ($z \lesssim 1$), while recover the $\Lambda$CDM universe at higher redshifts ($z \gg 1$) to be consistent with the standard cosmology. The cosmological parameters, $\Omega_{\rm m}$ and $H_0$, are determined by the CMB observation at higher redshifts. Then the luminosity distance $d_{\rm L}$ is mapped into redshift $z$ by
\begin{align}
d_{\rm L}(z) &= (1+z) \int_{0}^{z} \frac{dz^{\prime}}{H(z^{\prime})} \\
H(z) &= H_0 \left\{ \Omega_{\rm m} (1+z)^3 + (1-\Omega_{\rm m} ) \right\}^{1/2} \;. \nonumber
\end{align}
and $z$ is directly determined from GW observations.

In what follows, we classify modified-gravity waveform in Eq.~(\ref{eq9}) into two subclasses, $\nu\mu$ model with a redshift prior $\Delta z=10^{-3}$ and $\delta_{\rm g}\mu$ model with a timing prior $\Delta t_{\rm c}=1\,{\rm s}$, and consider them separately. This is because there are parameter degeneracies between $z$ and $\nu$ in $\nu\mu$ model and between $t_{\rm c}$ and $\delta_g$ in $\delta_{\rm g}\mu$ model, respectively. Since all dimensional quantities in the GW waveform, that is, masses and frequencies, are redshifted in the same way and degenerate with redshift, the redshift must be determined from a combination of
\begin{equation}
(1+z)^{-\nu/2} \frac{{\cal M}^{5/6}}{d_{\rm L}(z)} \;.
\end{equation}
The chirp mass is determined from GW phase, but $z$ and $\nu$ are completely degenerated. Therefore, we need source redshift information by identifying a host galaxy or detecting electromagnetic transient counterpart. Redshift information would be available even for BH binaries only if they are located at low redshift, $z<0.1$, and have high SNR or good angular resolution so that a unique host galaxy is identified \cite{Chen:2016tys,Nishizawa:2016ood}. On the other hand, from Eqs.~(\ref{eq9}), (\ref{eq:PhenomD-full}), and (\ref{eq:PNphase}), the quantity constrained from an observation in $\delta_{\rm g}\mu$ model is a combination of 
\begin{equation}
t_{\rm c} + \delta_g \frac{d_{\rm L}(z)}{1+z} \;. 
\label{eq:timing-combination}
\end{equation} 
To break the degeneracy and measure $\delta_g$ separately, we need to determine $t_{\rm c}$ from other observational means ($z$ is determined from GW amplitude). If a GW event is accompanied by an electromagnetic counterpart, $t_{\rm c}$ is estimated from difference of arrival times between a GW and an electromagnetic signal. Then $\delta_g$ is constrained in a certain range, depending on an uncertainty in $t_{\rm c}$ \cite{Nishizawa2014PRD}. 

To have an electromagnetic counterpart and obtain information about $t_{\rm c}$, we need NS-NS and NS- BH binary mergers, which are expected to accompany with some electromagnetic emissions \cite{Paschalidis:2016CQG,Rosswog:2016CQG}. For them, since we cannot apply the PhenomD waveform, we will use the inspiral waveform up to 3.5 PN order in phase by limiting the PhenomD waveform to
\begin{equation}
h_{\rm GR} = {\cal G}_I A_0 \, e^{i \phi_{\rm ins,E}} \quad \quad f \leq f_{\rm ISCO} \;,  
\end{equation}
with $A_0$ in Eq.~(\ref{eq:A0}) and $f_{\rm ISCO}=(6^{3/2}\pi)^{-1} f_M \approx 0.0217 f_M$, where $f_M \equiv M^{-1}$. Note that $f_{\rm ISCO}$ is twice the innermost stable circular orbit frequency for a point mass in Schwarzschild spacetime.

\subsection{Numerical setup}

In the following analysis, we will set fiducial parameters to $t_{\rm c}=\phi_c=\chi_s=\chi_a=\nu=\mu=\delta_g=0$ and randomly generate sky locations ($\theta_{\rm S}$, $\phi_{\rm S}$) and other angle parameters ($\iota$, $\psi$) for compact binaries with fixed masses and redshift. As for GW detectors, we consider a detector network composed of aLIGO at Hanford and Livingston, and aVIRGO (HLV), assuming they have the same noise curve as aLIGO \cite{Sathyaprakash:2009xs}. The signal-to-noise ratio (SNR) $\rho$ of each source is computed from
\begin{equation}
\rho^2 = 4 \sum_{I} \int_{f_{\rm{min}}}^{f_{\rm{max}}} \frac{|\tilde{h}_I (f)|^2}{S_h(f)} df \;,
\label{eq:SNR}
\end{equation}
where $\tilde{h}_I$ is the Fourier amplitude of a GW signal in $I$th detector and $S_{h}$ is the noise power spectral density of a detector. In the procedure of the source generation, we set the SNR threshold for detection and keep only sources with network SNR $\rho>8$.

The Fisher information matrix is given by \cite{Cutler:1994ys,Finn:1992wt}
\begin{equation}
\Gamma_{ab} = 4 \sum_{I} \, {\rm{Re}} \int_{f_{\rm{min}}}^{f_{\rm{max}}}
 \frac{\partial_{a} \tilde{h}_I^{\ast}(f)\, \partial_{b}
 \tilde{h}_I(f)}{S_{\rm{h}}(f)} df \;,
\end{equation}
where $\partial_a$ denotes a derivative with respect to a parameter $\theta_a$. To implement a Gaussian prior on $z$ and $t_{\rm c}$ in the Fisher matrix formalism, we add $1/(\Delta \log z)^2$ and $1/(\Delta t_{\rm c})^2$ to the $(\log z, \log z)$ and $(t_{\rm c}, t_{\rm c})$ components of the Fisher matrix, respectively. This is equivalent to multiplying a likelihood function by a prior probability distribution. We take a standard deviation of $z$ in $\nu\mu$ model as $\Delta z=0.001$ and $t_{\rm c}$ in $\delta_{\rm g}\mu$ model as $\Delta t_{\rm c}=1\,{\rm s}$. The choice of the $z$ prior is motivated by possible identification of a host galaxy with a spectroscopic observation (for NS-NS and NS- BH binary mergers, an electromagnetic transient counterpart is also expected), while the choice of the $t_{\rm c}$ prior is motivated by possible association of NS-NS and NS- BH binary mergers with short gamma-ray bursts and the estimation of arrival time difference between a GW and gamma-ray photons from consideration of the various emission mechanisms \cite{LiHu2016ApJ}. The parameter estimation errors are computed from the inverse Fisher matrix. We define the sky localization error as
\begin{equation}
\Delta \Omega_{\rm S} \equiv 2 \pi | \sin \theta_{\rm S}| \sqrt{(\Delta \theta_{\rm S})^2(\Delta \phi_{\rm S})^2 - \langle \delta \theta_{\rm S} \delta \phi_{\rm S} \rangle^2 } \;,
\end{equation}
where $\langle \cdots \rangle$ stands for ensemble average and $\Delta \theta_{\rm S} \equiv \langle (\delta \theta_{\rm S})^2 \rangle^{1/2}$ and $\Delta \phi_{\rm S} \equiv \langle (\delta \phi_{\rm S})^2 \rangle^{1/2}$.

\subsection{Results for $\nu\mu$ model}

\begin{figure*}[t]
\begin{center}
\raisebox{1.5mm}{\includegraphics[width=4.5cm]{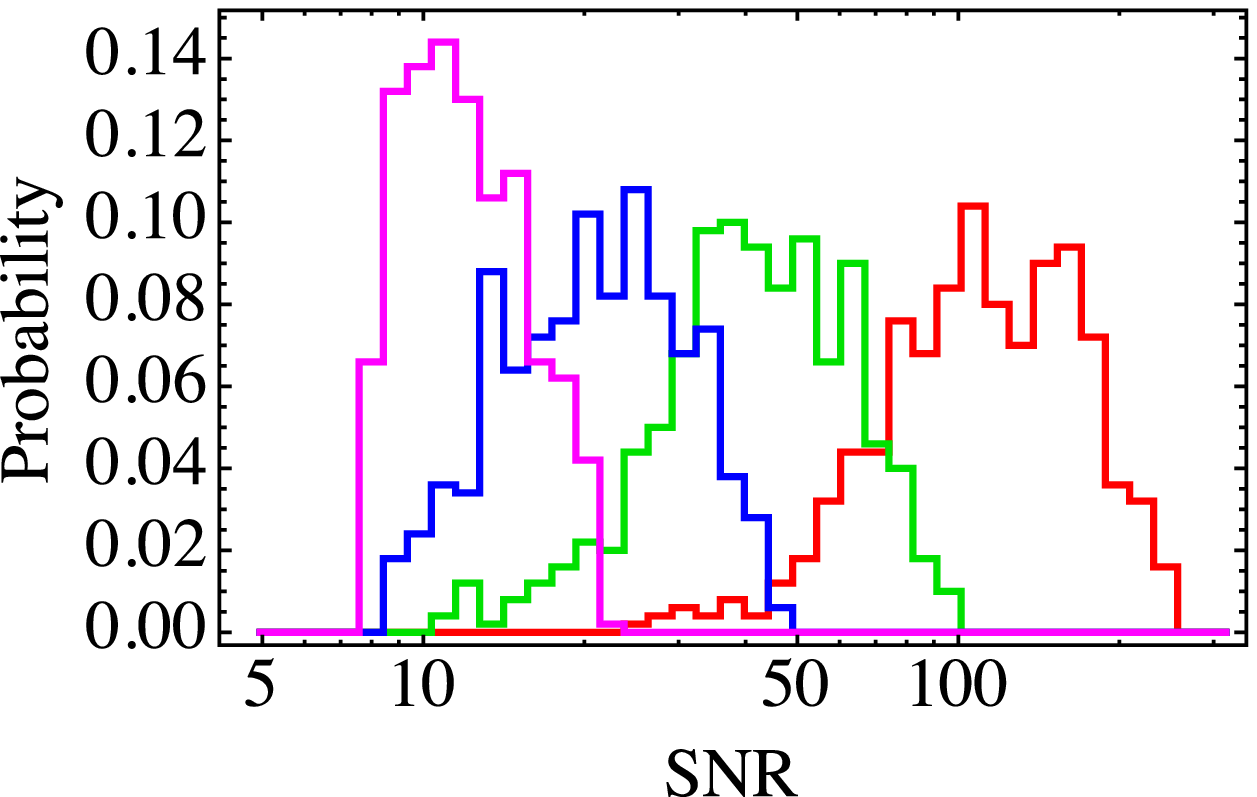}}
\hspace{2mm}
\raisebox{-2.2mm}{\includegraphics[width=4.5cm]{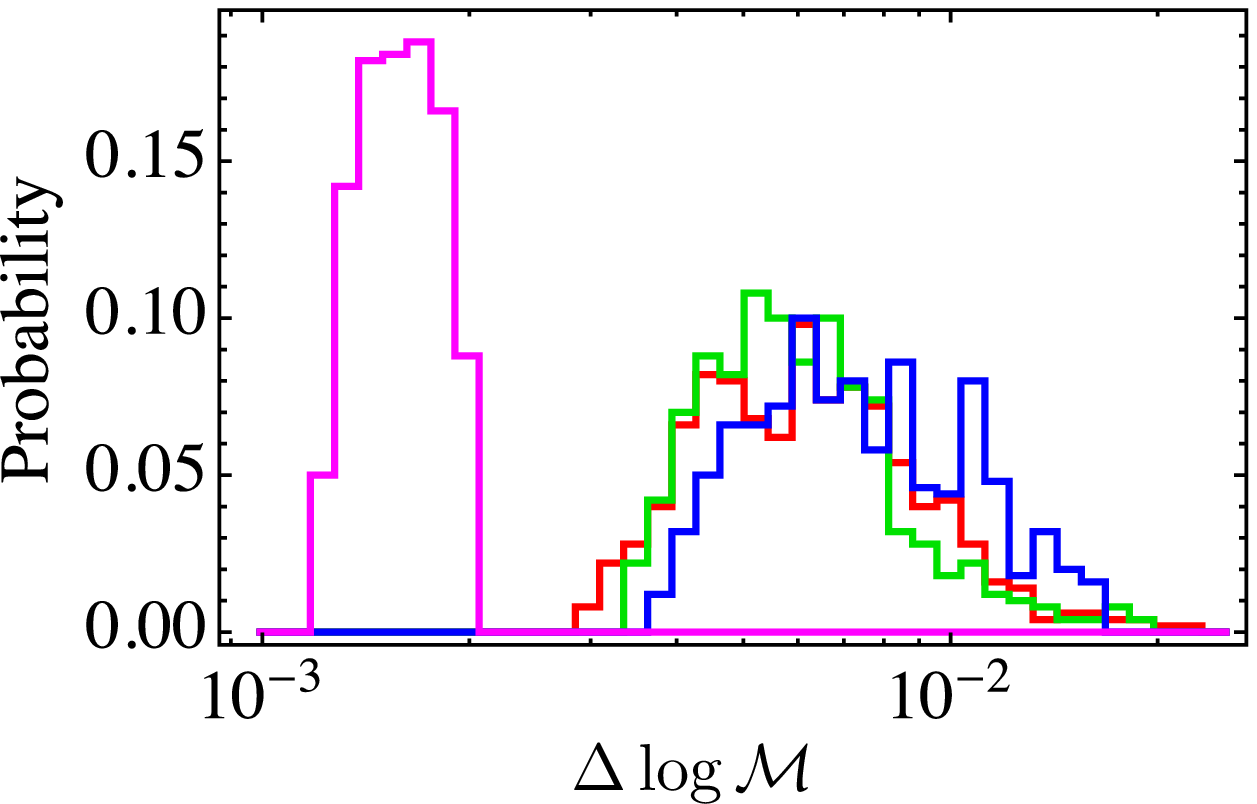}}
\hspace{2mm}
\raisebox{0mm}{\includegraphics[width=4.3cm]{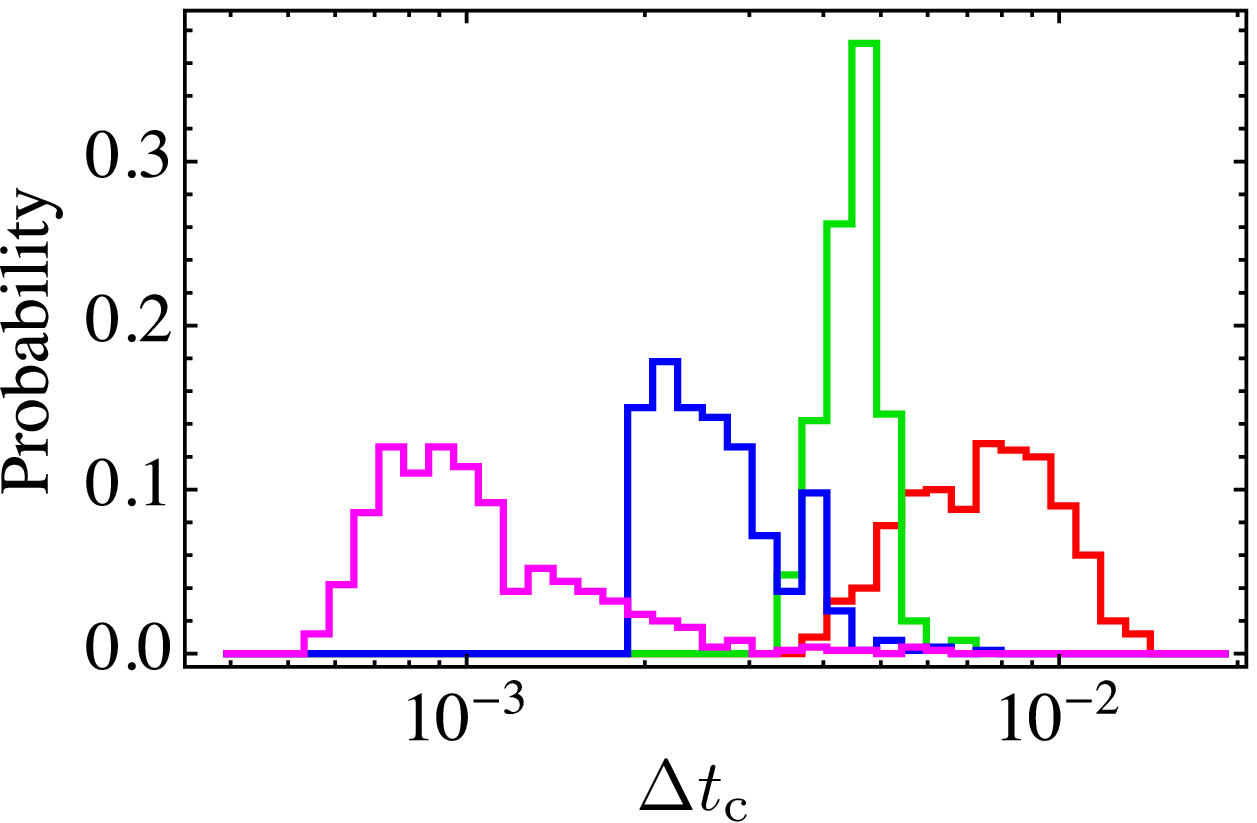}}
\raisebox{0mm}{\includegraphics[width=4.5cm]{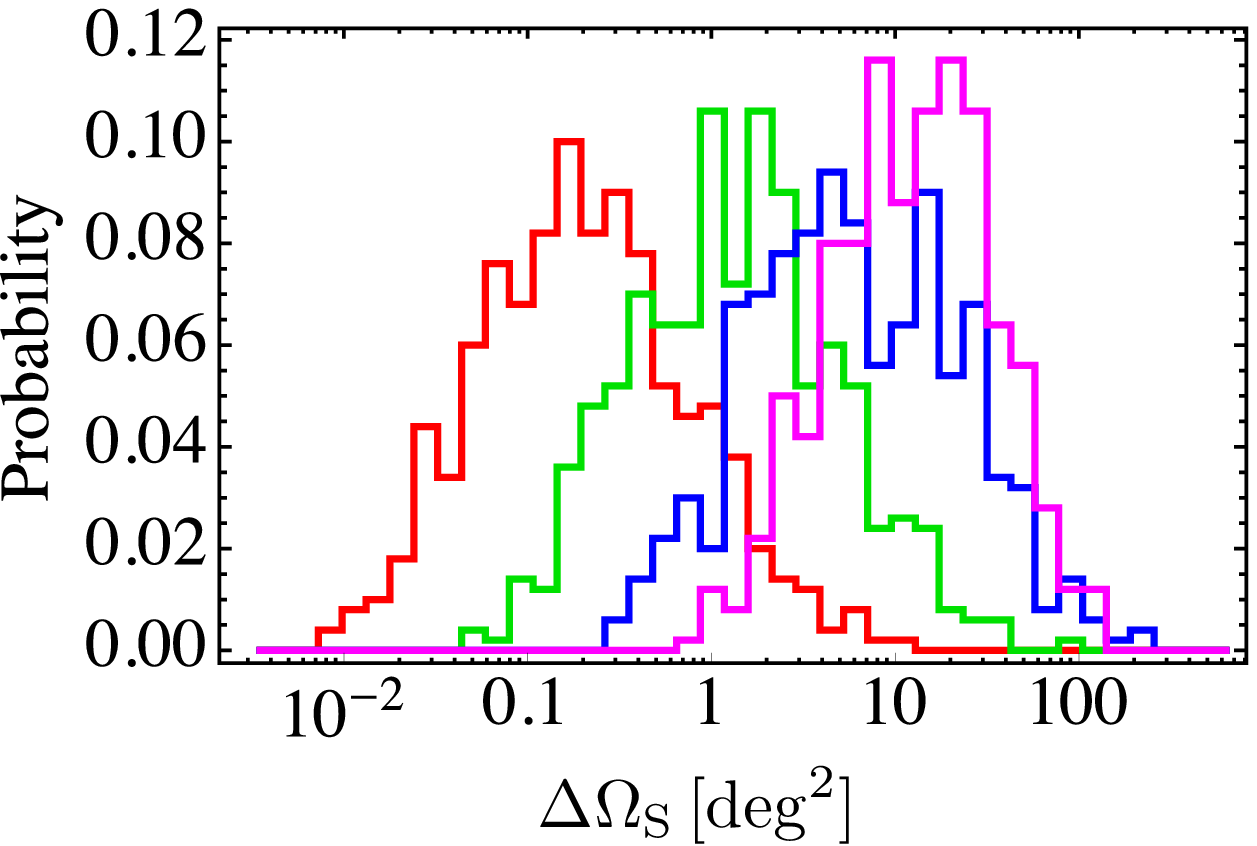}}
\hspace{2mm}
\raisebox{0.5mm}{\includegraphics[width=4.5cm]{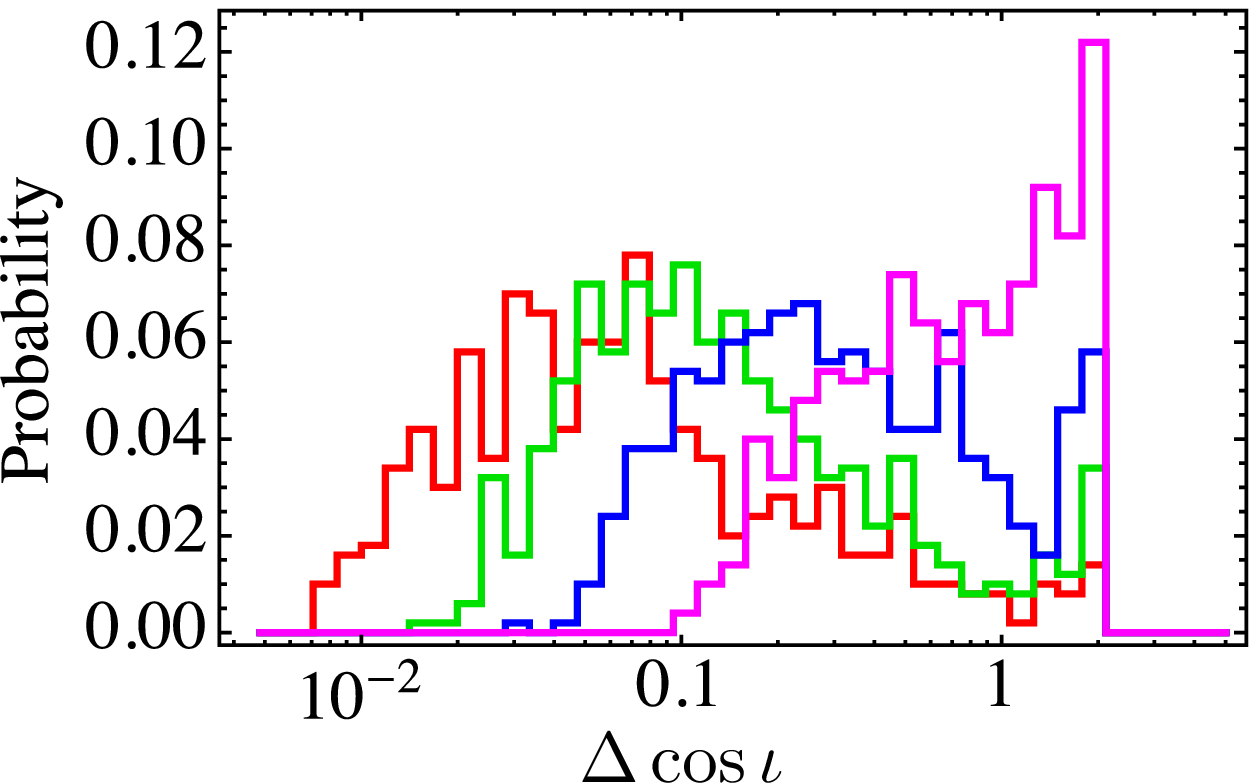}}
\hspace{2mm}
\raisebox{0.4mm}{\includegraphics[width=4.5cm]{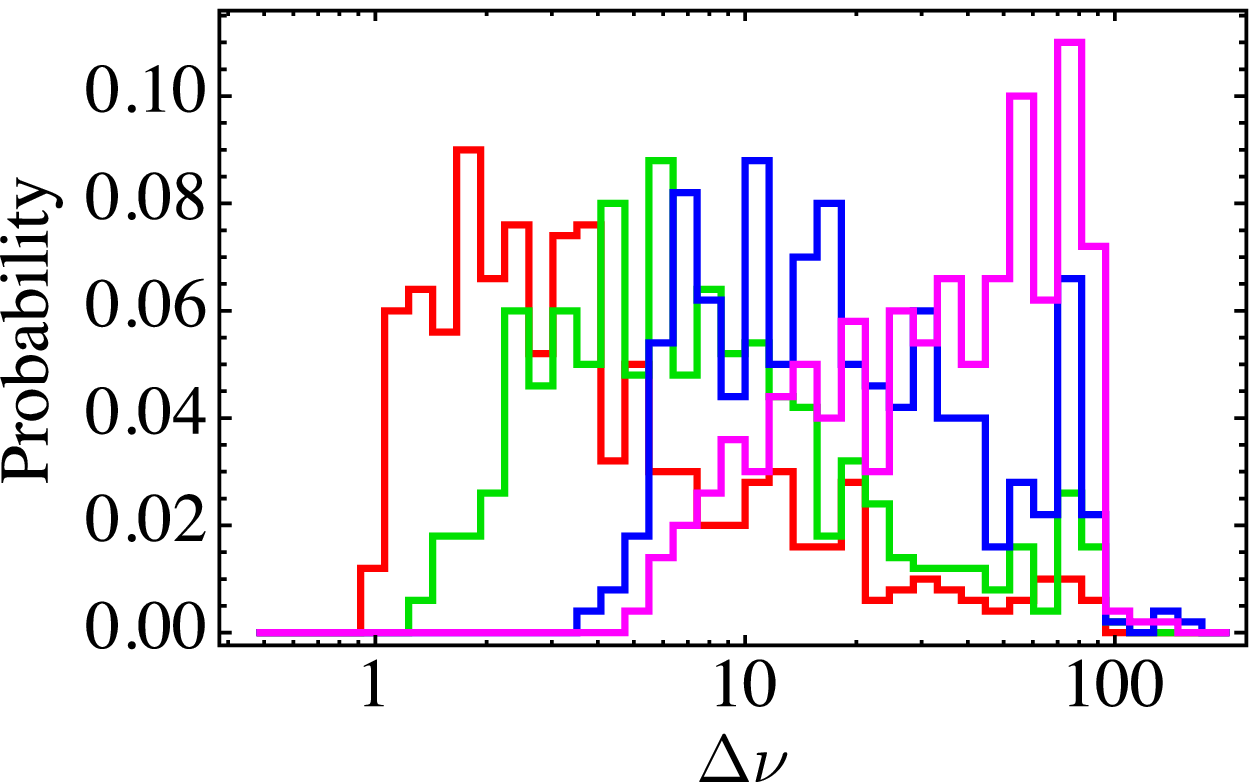}}
\hspace{4mm}
\raisebox{-0.7mm}{\includegraphics[width=4.5cm]{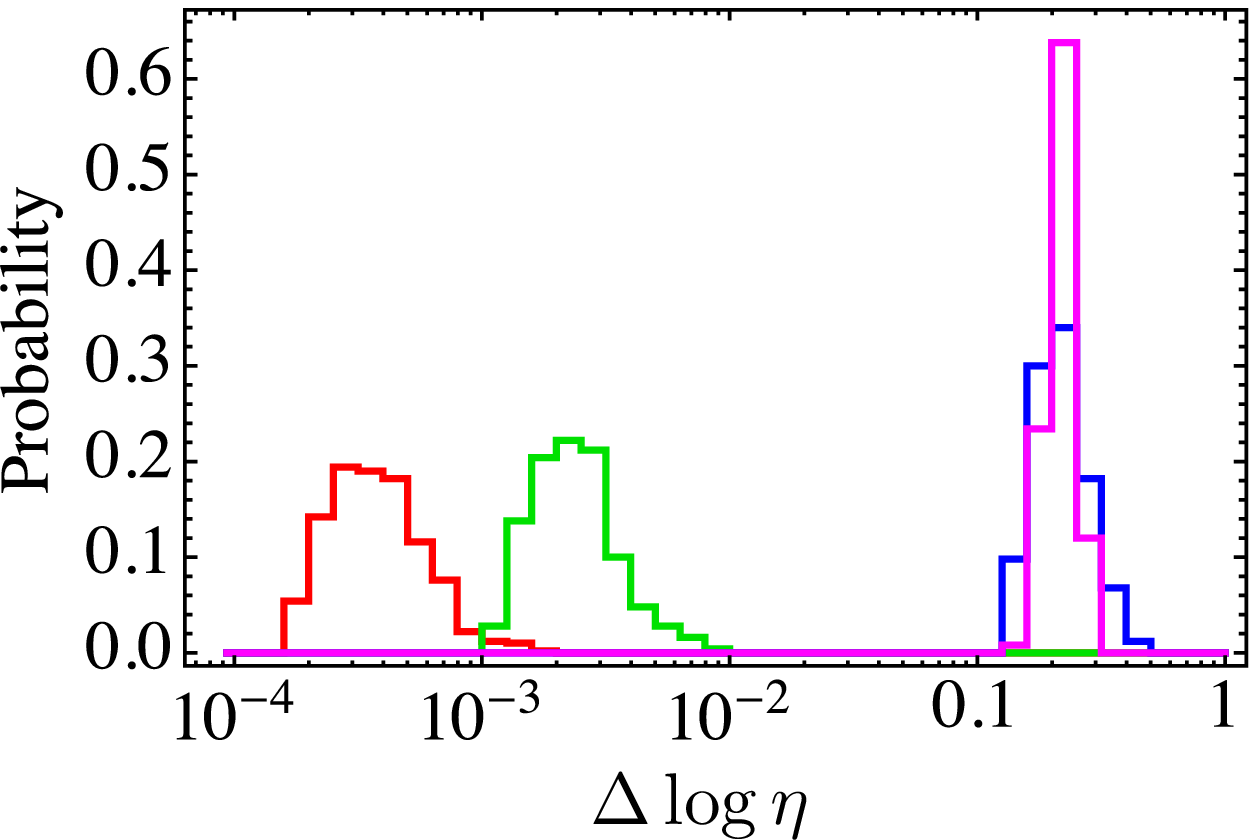}}
\hspace{3mm}
\raisebox{-0.3mm}{\includegraphics[width=4.5cm]{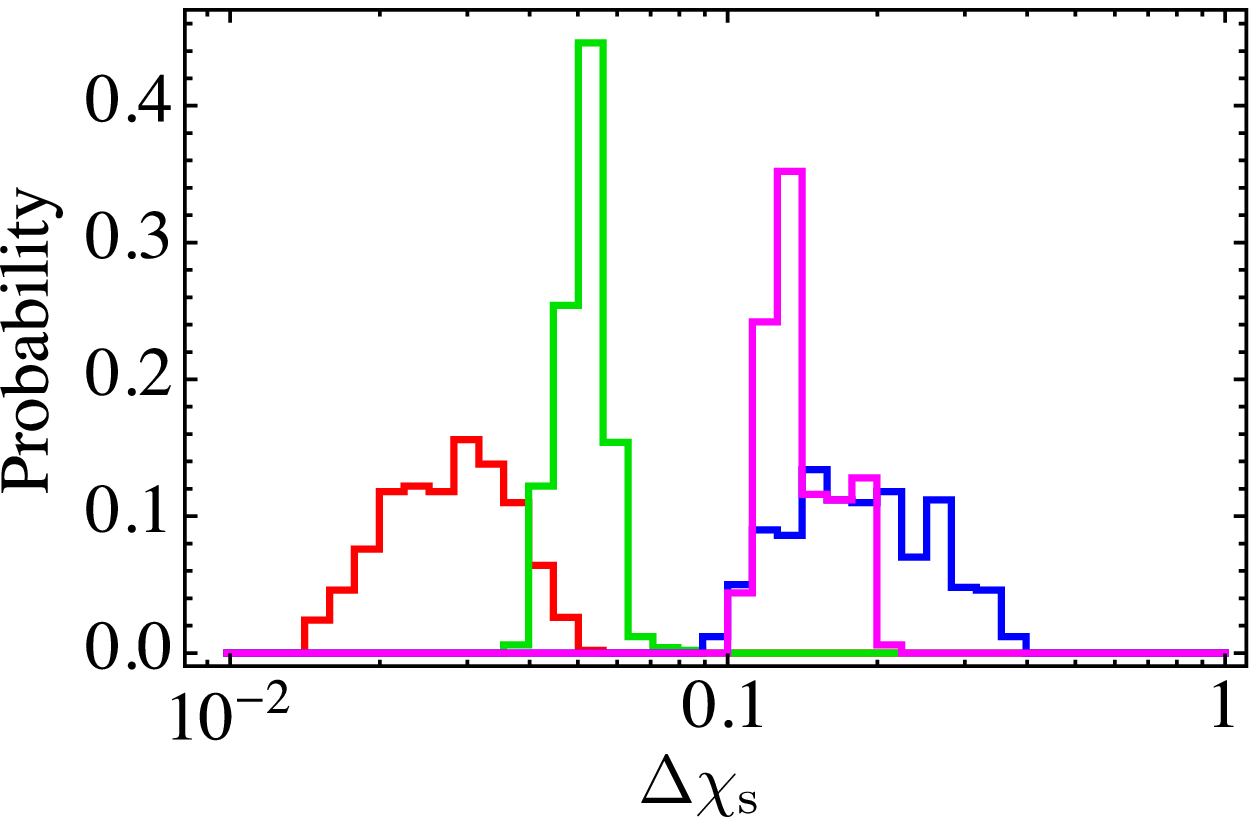}}
\hspace{-3mm}
\raisebox{-0.3mm}{\includegraphics[width=5.1cm]{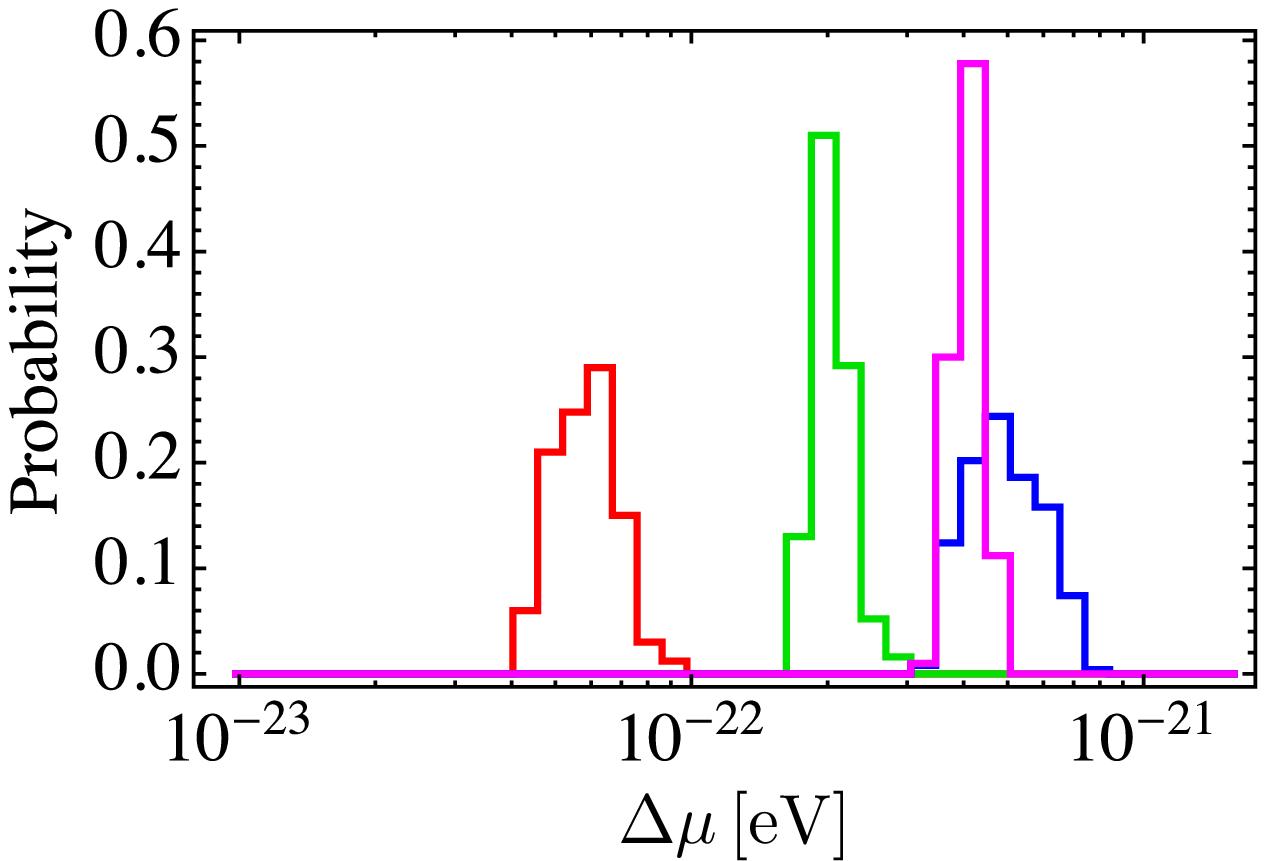}}
\caption{Parameter estimation errors in $\nu\mu$ model with a redshift prior, $\Delta z=10^{-3}$, showing mass dependence: $30M_{\odot}$-$30M_{\odot}$ (red), $10M_{\odot}$-$10M_{\odot}$ (green), $10M_{\odot}$-$1.4M_{\odot}$ (blue), $1.4M_{\odot}$-$1.4M_{\odot}$ (magenta). The redshift is fixed to $z=0.05$.}
\label{fig4}
\end{center}
\end{figure*}

We generated 500 sources for each class of compact binaries: $30M_{\odot}$BH -$30M_{\odot}$BH, $10M_{\odot}$BH - $10M_{\odot}$BH, $10M_{\odot}$BH -$1.4M_{\odot}$NS, and $1.4M_{\odot}$NS - $1.4M_{\odot}$NS, at $z=0.05$. As mentioned in the previous subsection, we add a redshift prior $\Delta z=10^{-3}$ to break the parameter degeneracy between $z$ and $\nu$. The results are shown in Fig.~\ref{fig4}. The larger chirp mass is, the larger SNR is. However, $\Delta \log {\cal M}$ is almost the same except for a $30M_{\odot}$ - $30M_{\odot}$ BH binary because ${\cal M}$ is highly correlated with $z$ and the error in $z$ is constrained by a prior $\Delta z=10^{-3}$. Only a $30M_{\odot}$ - $30M_{\odot}$ BH binary can determine ${\cal M}$ well below the prior width. On the other hand, $t_{\rm c}$ error is smaller for lighter binaries because their higher merger frequencies allow us to observe them longer and to determine phase parameters better. Other parameters, $\Omega_{\rm S}$, $\cos \iota$, $\eta$, $\chi_{\rm s}$, $\nu$, and $\mu$, basically trace the standard scaling, $\propto 1/{\rm SNR}$, though binaries with large mass-ratio are less sensitive to symmetric parameters with respect to component masses. The $\nu$ error distribution is similar to those of $\Omega_{\rm S}$ and $\cos \iota$ as they are correlated with $\nu$ though GW amplitude at Newtonian order. In other words, once the $z$ prior is imposed, these parameters scales with the standard SNR scaling and heavier binaries give smaller errors in $\nu$. While the $\mu$ error distribution is similar to $\log \eta$ and $\chi_{\rm s}$ because $\mu$ comes in the phase term at 1 PN order and is correlated with $\log \eta$ and $\chi_{\rm s}$ in the leading terms in phase at 1 PN and 1.5 PN orders, respectively.  Since the range of $\eta$ is limited to $\leq 0.25$, $\log \eta$ error has an upper limit. Consequently, the $\mu$ error of $1.4M_{\odot}$ - $1.4M_{\odot}$ NS binary cannot be so large.

\begin{figure*}[t]
\begin{center}
\raisebox{0mm}{\includegraphics[width=4.5cm]{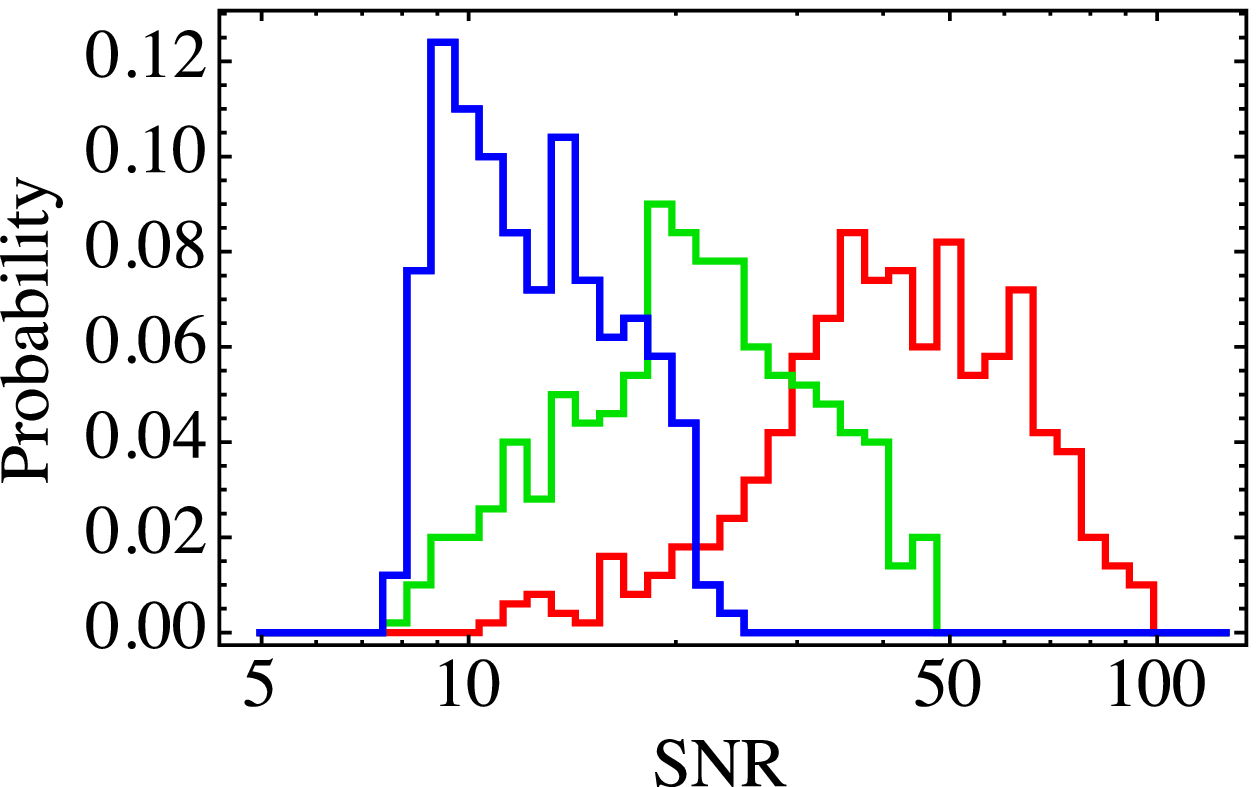}}
\hspace{2.5mm}
\raisebox{-1.4mm}{\includegraphics[width=4.5cm]{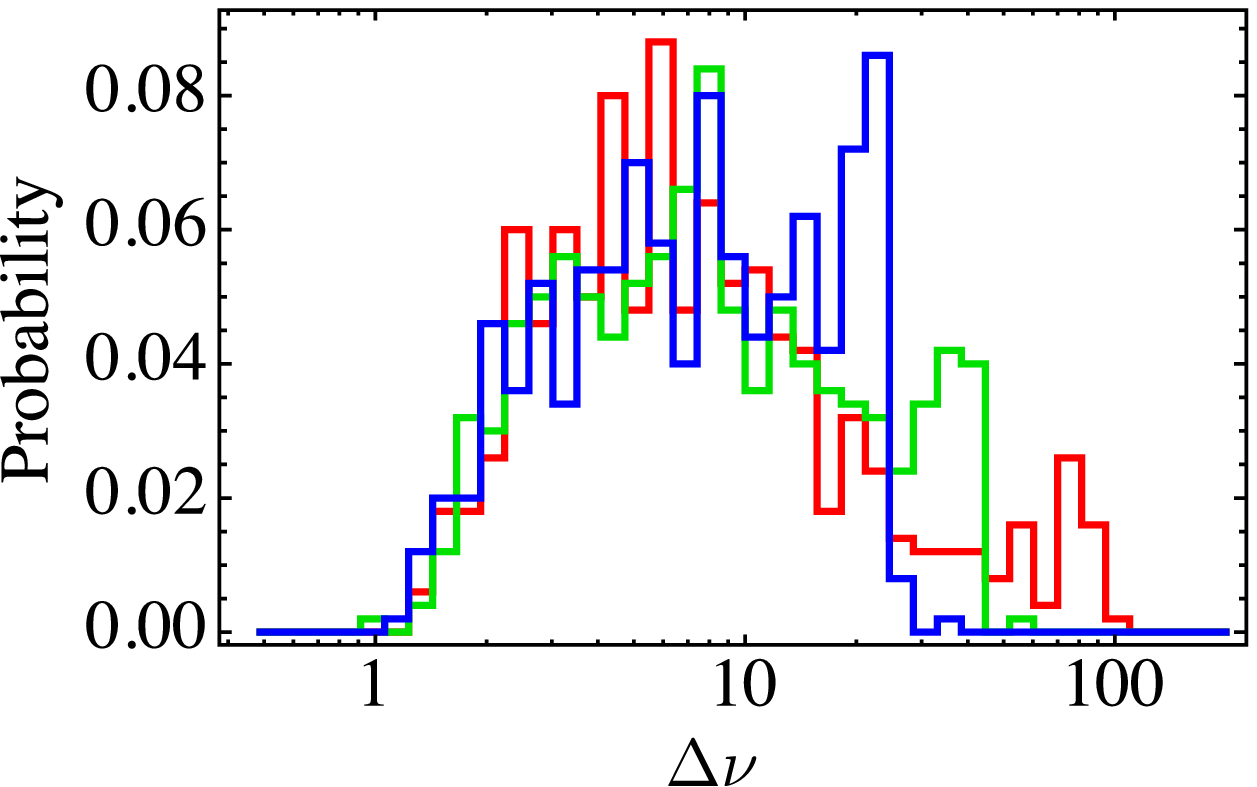}}
\hspace{2.5mm}
\raisebox{-1.2mm}{\includegraphics[width=4.5cm]{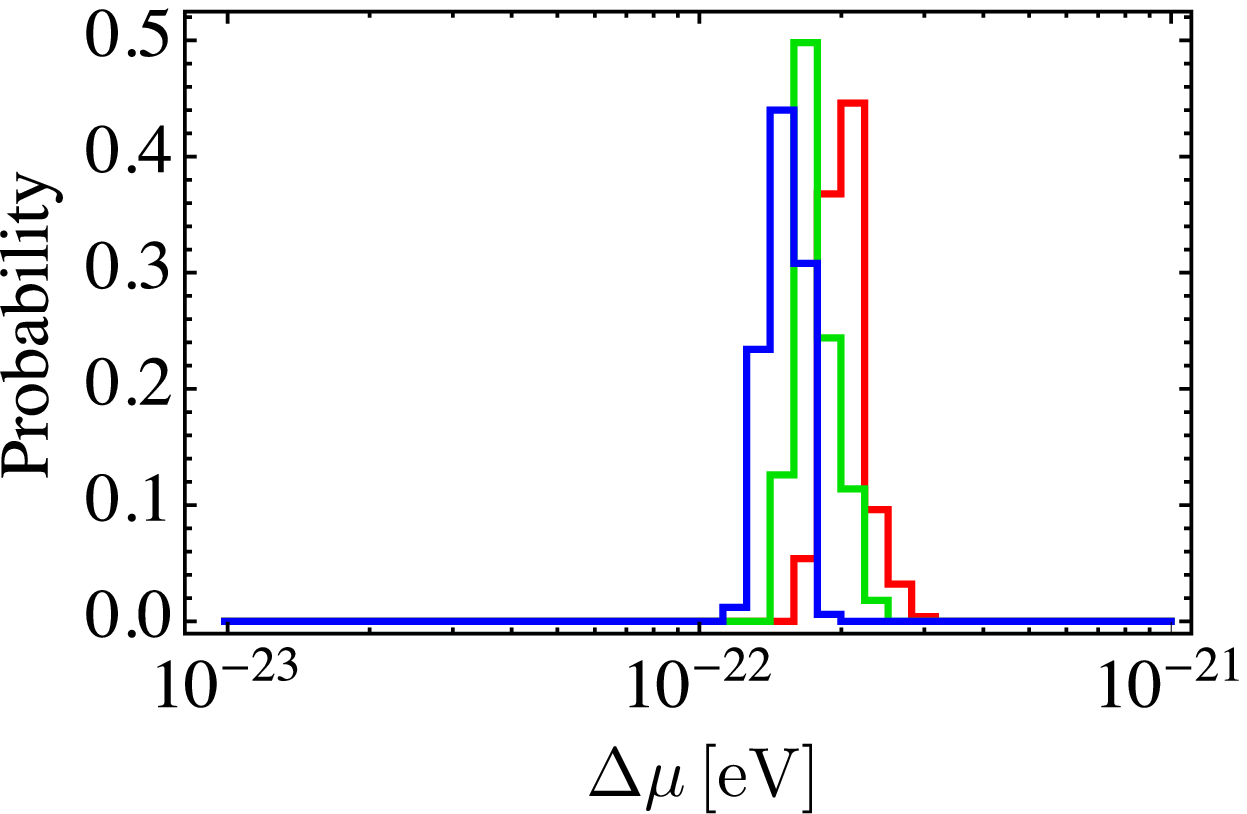}}
\caption{Parameter estimation errors $\nu\mu$ with a redshift prior, $\Delta z=10^{-3}$, showing redshift dependence: $z=0.05$ (red), $z=0.1$ (green), $z=0.2$ (blue). The masses are fixed to $10M_{\odot}$-$10M_{\odot}$.}
\label{fig5}
\end{center}
\end{figure*}

Figure~\ref{fig5} shows redshift dependence of $\nu$ and $\mu$ errors by generating 500 equal-mass BH binaries with $10\,M_{\odot}$ at $z=0.05$, $0.1$, and $0.2$. A remarkable feature is that the error distributions of $\nu$ and $\mu$ hardly depend on redshift. This is explained as follows. At low redshifts, SNR is inversely proportional to redshift and the parameter estimation errors become worse at far distance. On the other hand, the modified gravity effects are accumulated during propagation and become larger as distance increases. Then these scalings compensate each other and lead to the scaling almost independent of the source redshift. This indicates that sources at higher redshifts are likely to be used for constraining modified gravity parameters merely because they are more likely to be detected due to large comoving volume.   

\begin{table*}[t]
\begin{center}
\begin{tabular}{cccccc}
\hline \hline
$\;\;m_1\,[M_{\odot}]\;\;$ & $\;\;m_2\,[M_{\odot}]\;\;$ & $\;\;\Delta \nu$ (median) \;\; & $\;\;\Delta \nu$ (top 10\%) \;\; & $\;\;\Delta \mu \,[{\rm eV}]$ (median) \;\; & $\;\;\Delta \mu\,[{\rm eV}]$ (top 10\%) \;\; \\
\hline  
30 & 30 & 3.21 & 1.33 & $5.85\times 10^{-23}$ & $4.74\times 10^{-23}$ \\
10 & 10 & 6.37 & 2.46 & $2.03\times 10^{-22}$ & $1.82\times 10^{-22}$ \\
10 & 1.4 & 16.1 & 6.54 & $4.89\times 10^{-22}$ & $3.87\times 10^{-22}$ \\
1.4 & 1.4 & 35.9 & 9.92 & $4.09\times 10^{-22}$ & $3.71\times 10^{-22}$ \\
\hline \hline 
\end{tabular}
\end{center}
\caption{Median and top 10\% errors of parameter estimation in $\nu\mu$ model when the redshift is fixed to $z=0.05$.}
\label{tab3}
\end{table*}

In Table~\ref{tab3}, the errors in $\nu$ and $\mu$ are summarized. In conclusion, with the help of the $z$ prior, we can achieve the measurement of $\nu$ up to at a level of $\Delta \nu \approx 1.3$ by observing a single source.

\subsection{Results for $\delta_{\rm g}\mu$ model}

\begin{figure*}[t]
\begin{center}
\raisebox{0mm}{\includegraphics[width=4.5cm]{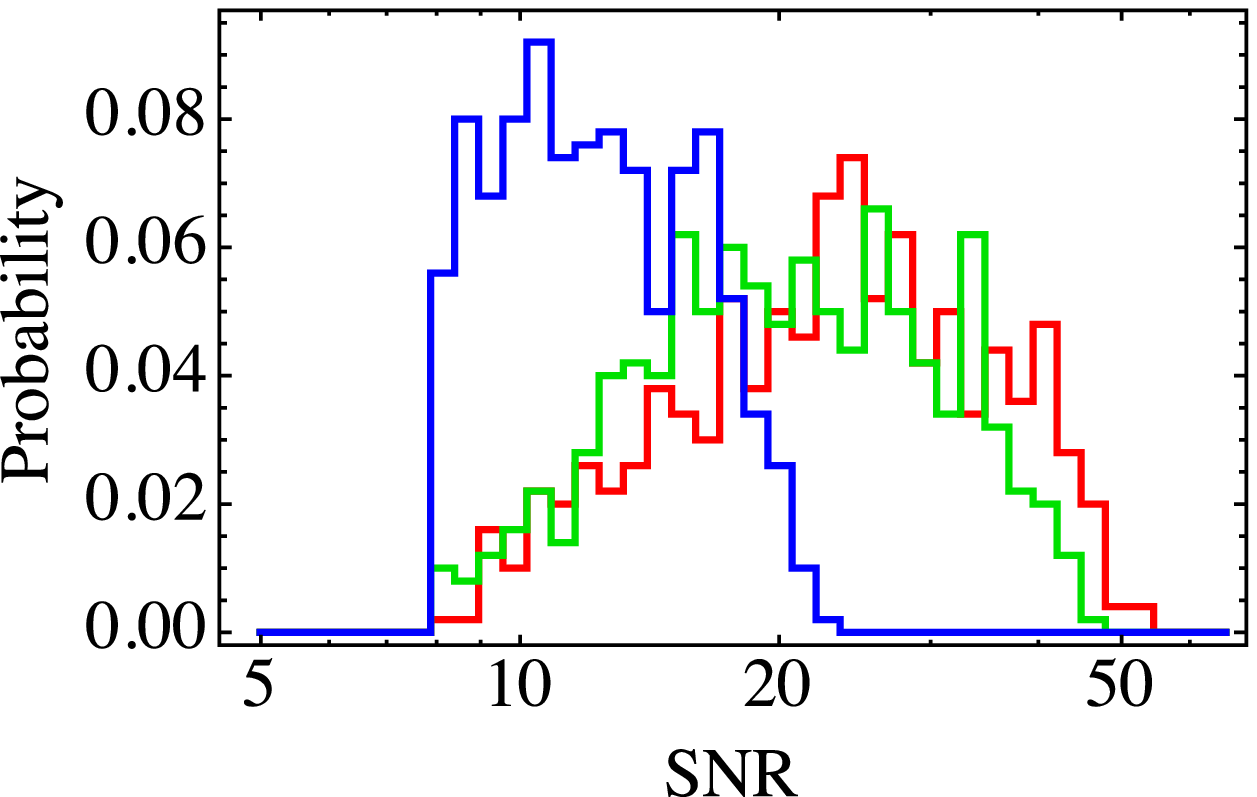}}
\hspace{2mm}
\raisebox{-4mm}{\includegraphics[width=4.5cm]{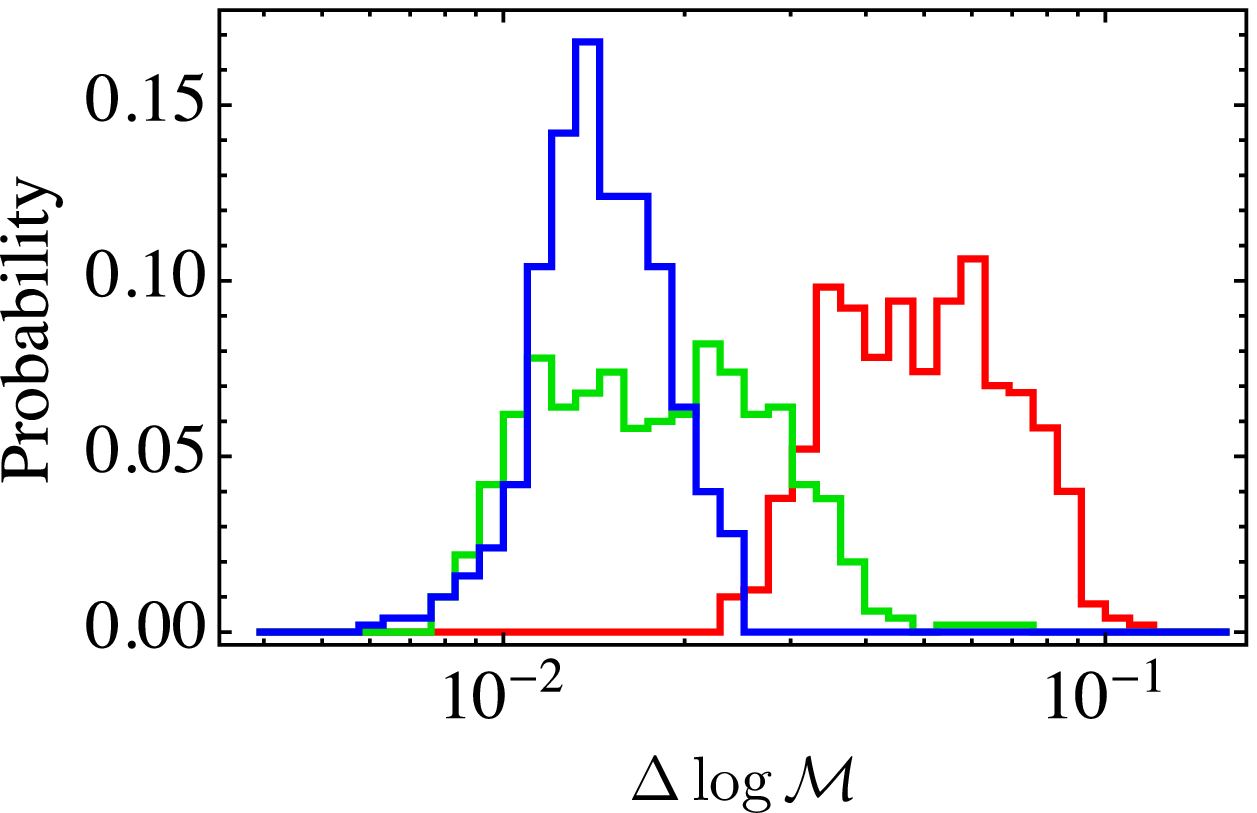}}
\hspace{2mm}
\raisebox{-0.2mm}{\includegraphics[width=4.5cm]{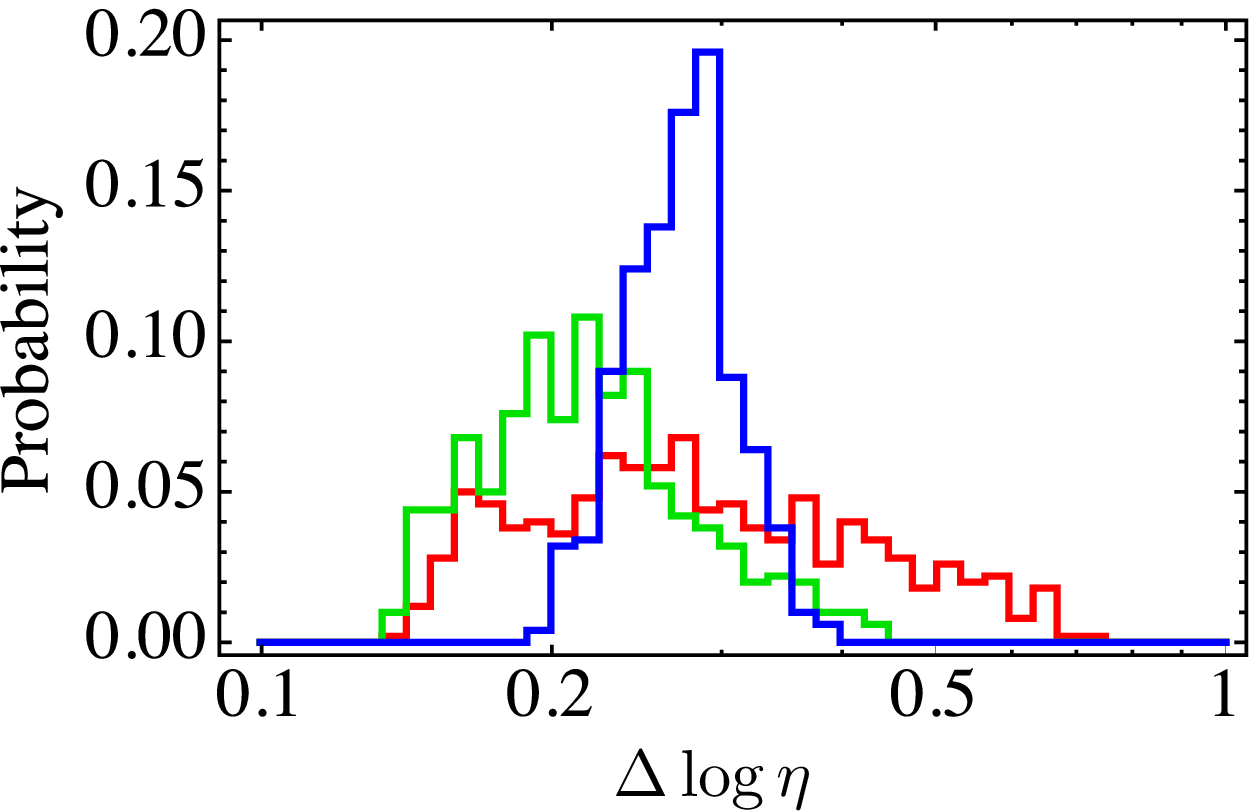}}
\raisebox{2.2mm}{\includegraphics[width=4.5cm]{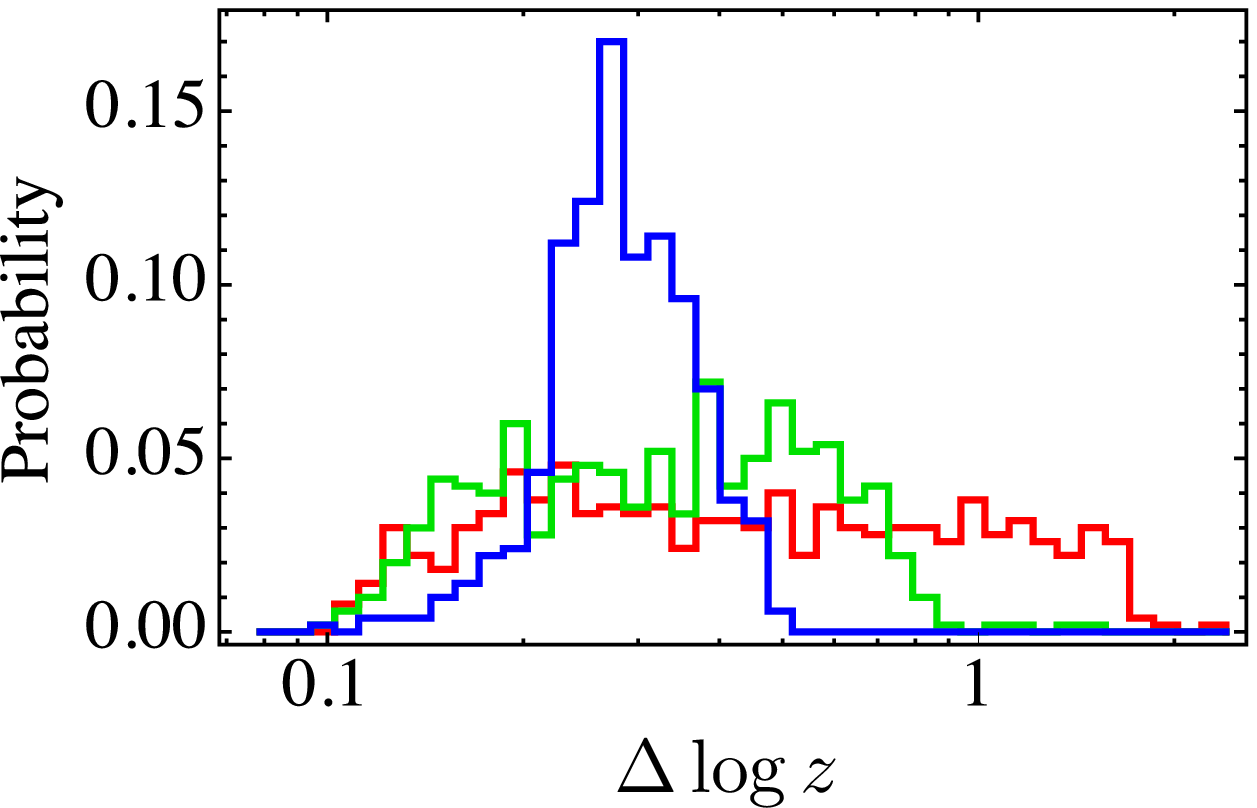}}
\hspace{2mm}
\raisebox{-0.8mm}{\includegraphics[width=4.5cm]{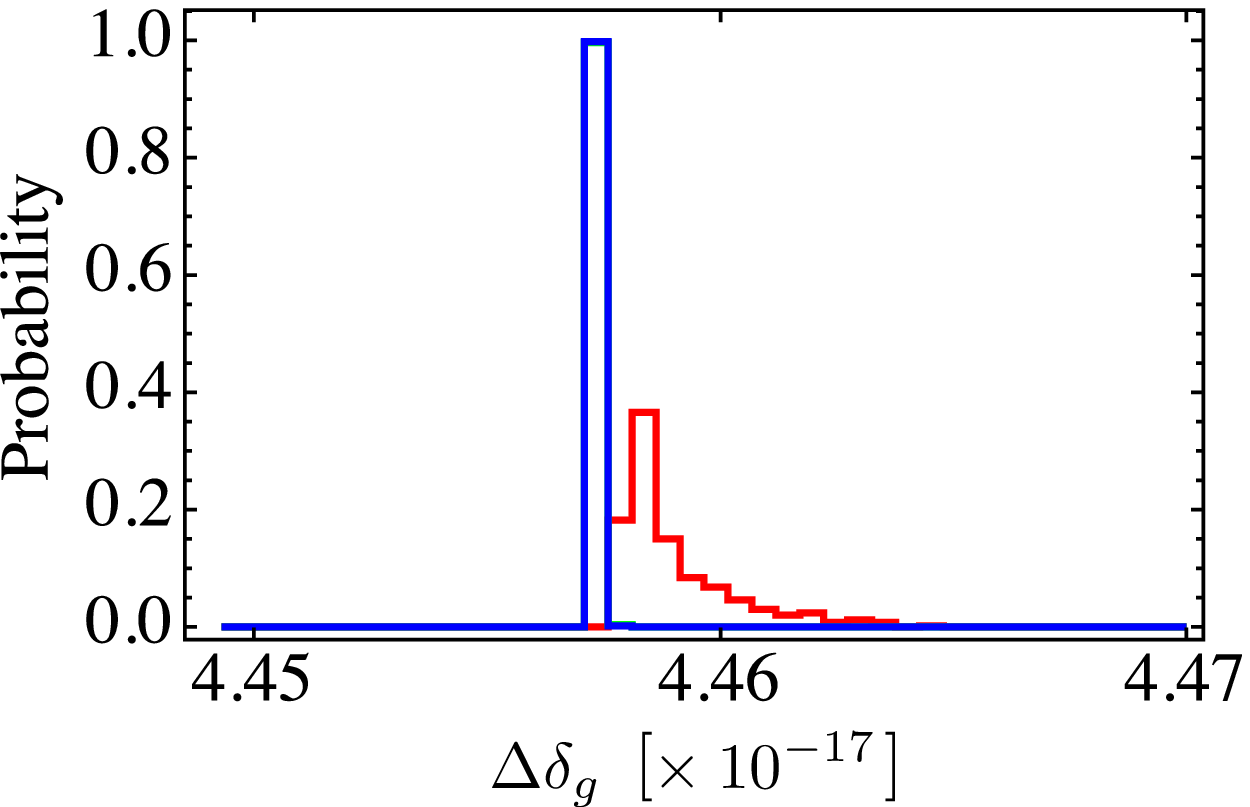}}
\hspace{2mm}
\raisebox{1.8mm}{\includegraphics[width=4.5cm]{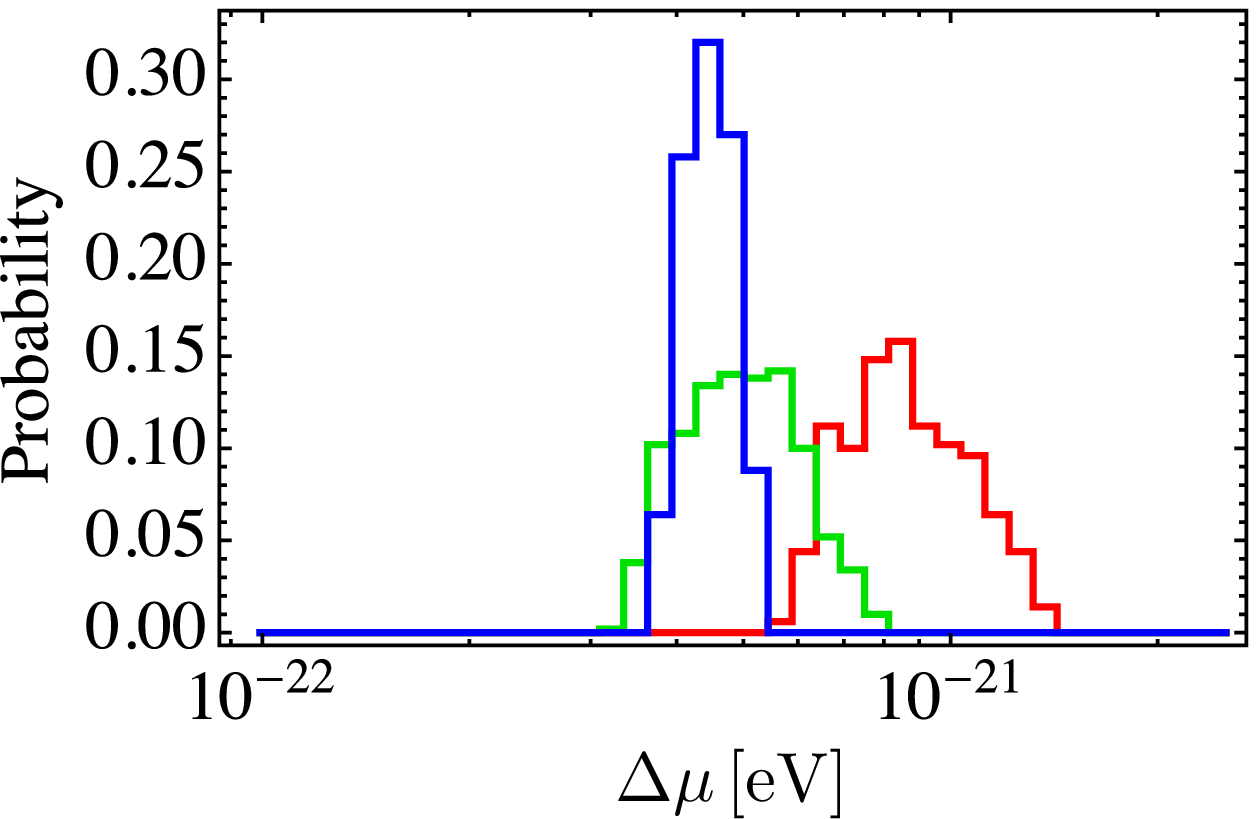}}
\caption{Parameter estimation errors in $\delta_g\mu$ model with $t_{\rm c}$ prior, $\Delta t_{\rm c}=1\,{\rm s}$, showing mass dependence: $30M_{\odot}$-$1.4M_{\odot}$ (red), $10M_{\odot}$-$1.4M_{\odot}$ (green), $1.4M_{\odot}$-$1.4M_{\odot}$ (blue). The redshift is fixed to $z=0.05$. In the $\delta_g$ plot, the green line is completely overlapped with the blue line.}
\label{fig1}
\end{center}
\end{figure*}

We generated 500 sources for each class of compact binaries: $30M_{\odot}$BH -$1.4M_{\odot}$NS, $10M_{\odot}$BH - $1.4M_{\odot}$NS, and $1.4M_{\odot}$NS -$1.4M_{\odot}$NS, at $z=0.05$. As we mentioned in Sec.~\ref{sec4a}, $t_{\rm c}$ and $\delta_g$ are completely degenerated. To break the degeneracy, we impose $t_{\rm c}$ prior, $\Delta t_{\rm c}=1\,{\rm s}$, assuming an electromagnetic counterpart. The results are shown in Fig.~\ref{fig1}. The dependences of the parameter estimation errors are much more complicated in $\delta_{\rm g}\mu$ model than in $\nu\mu$ model because of different mass ratios. The SNR of $30M_{\odot}$ -$1.4M_{\odot}$ and $10M_{\odot}$ - $1.4M_{\odot}$ binaries are almost same, but the mass ratio is different by three times, leading to different durations of an inspiral phase. That is why $10M_{\odot}$ - $1.4M_{\odot}$ binary can better determine mass parameters, ${\cal M}$ and $\eta$, and graviton mass $\mu$. The error of $\delta_{\rm g}$ is exactly the same for all binaries because this is constrained merely by the $t_{\rm c}$ prior.  

We also studied the redshift dependence of $\delta_{\rm g}$ error by generating 500 $10M_{\odot}$BH - $1.4M_{\odot}$NS binaries at $z=0.05$, $0.1$, and $0.2$. The interesting feature is that $\delta_g$ is well determined at high redshifts, in contrast to $\nu$ and $\mu$ errors. This is because in Eq.~(\ref{eq:timing-combination}), the quantity constrained by the $t_{\rm c}$ prior is 
\begin{equation}
\delta_{\rm g} \frac{d_{\rm L}(z)}{1+z} \;.
\end{equation} 
Since this term is roughly proportional to $\delta_{\rm g} z$ at low redshifts and is constrained to be $\lesssim 1\,{\rm s}$, then $\delta_g$ is better constrained at higher redshifts, irrespective of SNR. Indeed, as shown in Table~\ref{tab4}, for $10M_{\odot}$ BH -$1.4M_{\odot}$ NS binaries at $z=0.05$, $0.1$, and $0.2$, $\delta_g$ error scales as $4.5 \times 10^{-17}$, $2.3 \times 10^{-17}$, $1.2 \times 10^{-18}$ as the redshift increases, though their median SNR are 20.5, 11.9, 9.1, respectively. Therefore, in the $\delta_{\rm g}\mu$ model, the $t_{\rm c}$ prior plays an essential role to determine the parameter estimation precision of $\delta_g$, while $\mu$ error weakly depends on a source redshift.

\begin{table*}[t]
\begin{center}
\begin{tabular}{ccccccc}
\hline \hline
$\;\;m_1\,[M_{\odot}]\;\;$ & $\;\;m_2\,[M_{\odot}]\;\;$ & $\;\;\;z\;\;\;$ & $\;\;\Delta \delta_{\rm g}$ (median) \;\; & $\;\;\Delta \delta_{\rm g}$ (top 10\%) \;\; & $\;\;\Delta \mu \,[{\rm eV}]$ (median) \;\; & $\;\;\Delta \mu\,[{\rm eV}]$ (top 10\%) \;\; \\
\hline  
30 & 1.4 & 0.05 & $4.46\times 10^{-17}$ & $4.46\times 10^{-17}$ & $8.53\times 10^{-22}$ & $6.58\times 10^{-22}$ \\
10 & 1.4 & 0.05 & $4.46\times 10^{-17}$ & $4.46\times 10^{-17}$ & $4.95\times 10^{-22}$ & $3.86\times 10^{-22}$ \\
1.4 & 1.4 & 0.05 & $4.46\times 10^{-17}$ & $4.46\times 10^{-17}$ & $4.43\times 10^{-22}$ & $4.00\times 10^{-22}$ \\ 
10 & 1.4 & 0.1 & $2.26\times 10^{-17}$ & $2.26\times 10^{-17}$ & $4.77\times 10^{-22}$ & $3.94\times 10^{-22}$ \\
10 & 1.4 & 0.2 & $1.16\times 10^{-17}$ & $1.16\times 10^{-17}$ & $4.19\times 10^{-22}$ & $3.89\times 10^{-22}$ \\
\hline \hline 
\end{tabular}
\end{center}
\caption{Median and top 10\% errors of parameter estimation in $\delta_g\mu$ model.}
\label{tab4}
\end{table*}

\section{Current constraints and future prospect}
\label{sec5}

\subsection{Graviton mass $\mu$}

Currently graviton mass has been constrained by several observations of the galaxy, the solar system, and binary pulsars (for summary, see \cite{Berti:2011jz} and references therein). However, the constraints from the galaxy and the solar system have been obtained from the observations in static gravitational fields and cannot be applied directly to GWs. The only mass limit from dynamical gravitational fields had been that from binary pulsars for a long time: $m_g < 7.6\times 10^{-20}\,{\rm{eV}}$ \cite{Finn:2001qi}. Recently, aLIGO have detected gravitational waves from BH binaries and updated the dynamical mass bound, combining three GW events: $m_g < 7.7\times 10^{-23}\,{\rm{eV}}$ \cite{GW170104:detection}. This is close to our best forecast for the constraint on graviton mass in the case of a $30M_{\odot}$ - $30M_{\odot}$ BH binary at $z=0.05$ in the $\nu\mu$ model ($m_g < 4.7\times 10^{-23}\,{\rm{eV}}$). Therefore, there is no room for significant improvement of the mass constraint in aLIGO era. 

As expected from Eq.~(\ref{eq:time-delay}), graviton mass bound can be tighter at lower frequencies. There have been proposals for the possible constraints on graviton mass from the future observation of a compact binary with a space-based GW detector such as LISA \cite{Audley:2017drz} in the millihertz band and DECIGO \cite{Sato:2017dkf} in the decihertz band. By observing $10^7M_{\odot}$ - $10^6M_{\odot}$ BH binary at 3 Gpc with LISA, one can impose a limit $m_g < 4.0\times 10^{-26}\,{\rm{eV}}$ \cite{Yagi:2009zm}, while observing $10^6M_{\odot}$ - $10^5M_{\odot}$ BH binary at 3 Gpc with DECIGO gives a limit $m_g < 3.7\times 10^{-25}\,{\rm{eV}}$ \cite{Yagi:2009zz}. These constraints are about $10^2$ - $10^3$ times stronger than the aLIGO bound.

\subsection{Propagation speed $c_{\rm T}$}

GW propagation speed has been constrained indirectly from ultra-high energy cosmic rays. Assuming the cosmic rays originate in our Galaxy (conservatively assuming short propagation distance), the absence of gravitational Cherenkov radiation and the consequent observation of such cosmic rays on the Earth lead to the limit on GW speed, $\delta_g < 2 \times 10^{-15}$ \cite{Moore:2001bv}. However, this constraint on GW propagation speed (phase velocity) can be applied only to a subluminal case at very high energy $\sim 10^{10}\,{\rm GeV}$ or very high frequency $\sim 10^{33}\,{\rm Hz}$. While from the observational data of the orbital decay of a binary pulsar, the constraint on GW speed has been obtained, limiting superluminal propagation: $|\delta_g| \lesssim 10^{-2}$ \cite{Jimenez:2015bwa}. On the other hand, the first three detections of GW from BH binaries allow us for the first time to directly measure GW speed on the Earth, based on arrival time difference between detectors. Cornish {\it et al.} \cite{Cornish:2017jml} has given a new constraint on GW group velocity, $-0.42 < 1-v_{\rm g} < 0.45$, by combining the first three GW events in a Bayesian analysis with a linear prior on $v_{\rm g}$. Assuming $v_{\rm g}$ is constant, one can convert the constraint on $v_{\rm g}$ into $-0.42 < \delta_g < 0.45$.
This constraint is rather weak, but robust and reliable. More importantly, this is obtained in the high-density and relatively strong-gravity environment on the Earth, where the screening effect of modified gravity such as the chameleon mechanism \cite{Khoury:2003rn} and the Vainshtein mechanism \cite{Vainshtein:1972sx} may work.


Recently, a coincidence event between GW from a NS binary merger and a short gamma-ray burst, GW170817/GRB170817A, was detected \cite{GW170817:detection}. Assuming the emission of the gamma ray is not delayed more than $10\,{\rm sec}$ from that of GW and using the observed difference of the arrival times $1.7\,{\rm sec}$ and conservative distance to the source $d_{\rm L}=26\,{\rm Mpc}$ \cite{GW170817GRB}, the constant propagation speed of GW is constrained tightly so that $-7 \times 10^{-16} < \delta_g < 3 \times 10^{-15}$. This is consistent with our best forecast for the bound on GW speed $|\delta_ g| < 1.2 \times 10^{-17}$, because we assume a $t_{\rm c}$ prior $\Delta t_{\rm c}=1\,{\rm s}$ and a GW source at $z=0.2$ ($\sim 1\,{\rm Gpc}$), which are a slightly better prior and much larger distance. From the theoretical point of view in modified gravity theories, the GW propagation speed is not always constant but is likely to evolve with time. The constraints in time-dependent cases are discussed in detail in the subsequent paper of this series \cite{Arai:2017hxj}.  





\subsection{Amplitude damping rate $\nu$}

The amplitude damping rate $\nu$ has not yet been constrained well. Our best forecast for the constraint on $\nu$ is obtained from a $30M_{\odot}$ - $30M_{\odot}$ BH binary at $z=0.05$ to be $|\nu| < 1.3$, only if the source redshift is obtained by the identification and spectroscopic observation of a host galaxy. However, it is not easy to identify a host galaxy with an aLIGO-like detector network because of poor angular resolution. However, the very small number of GW events with redshift information would be obtain with aLIGO-like detector network at design sensitivity \cite{Chen:2016tys,Nishizawa:2016ood}. Using multiple BH-BH binaries in a few-year observation, $\nu$ would be able to be measured at the order of ${\cal O} (0.1)$. While the constraints from BH-NS or NS-NS binaries are much weaker than that from a BH-BH binary. Indeed the recent detection of GW170817 was accompanied with electromagnetic emissions in the broad range of frequencies and the redshift of the host galaxy was identified successfully. However, the constraint from GW170817 is too weak to test realistic models of modified gravity ($\Delta \nu \approx 80$) \cite{Arai:2017hxj}. The rate of such a event is still largely uncertain, but if a number of GW events with source redshifts is available, the constraint can be improved statistically by using multiple sources. Then the bound can be comparable to that from a single $30M_{\odot}$ - $30M_{\odot}$ BH-BH binary if 30 BH-NS binaries or 60 NS-NS binaries are detected with any electromagnetic counterparts.

One of other methods proposed so far to measure GW amplitude damping is the number count of GW sources \cite{Calabrese:2016bnu}. Once it is assumed that a binary merger rate is constant, the power index of GW amplitude damping, $d_{\rm L}^{-\gamma}$, is determined from a source number distribution in distance. According to \cite{Calabrese:2016bnu}, $\gamma$ is measured at 15\% precision with 100 sources observed by aLIGO under the assumption that binary parameters are completely known. Since the assumptions on the merger rate and the binary parameters are too strong in practice, it is difficult to compare with our result. But, since $\Delta \nu \sim \Delta \gamma$ at the leading order, the naive correspondence leads to the measurement of $\nu$ with an error of 0.15. Further study is necessary to conclude which method is better in realistic conditions.

\section{Conclusion}
\label{sec6}

To treat tests of gravity with GW more exhaustively and intuitively, irrespective of the models of gravity theories, GW sources, and background spacetimes, we have proposed a new universal framework for testing gravity, based on the propagation equation of a GW in an effective field theory. By analytically solving the GW propagation equation, we obtained a WKB solution with arbitrary functions of time that describe modified amplitude damping, modified propagation speed, nonzero graviton mass, and a possible source term for a GW. Then we have performed a parameter estimation study with the Fisher information matrix, showing how well the future observation of GW can constrain the model parameters in generalized models of GW propagation. One of the advantages to consider GW propagation is that even if modification on gravity is a tiny effect, propagation from a distant source can accumulate the effect and amplify a signal observed at a detector. 

For the constant $\nu\mu$ model, since $\nu$ and $z$ are completely degenerated, we need to impose a prior on redshift. Once the redshift information is obtained from the spectroscopic observation of a host galaxy or an electromagnetic transient, $\nu$ can be determined at a precision of $\Delta \nu \sim 1.3$ by observing a $30M_{\odot}$ - $30M_{\odot}$ BH binary at $z=0.05$. While our best forecast for the constraint on graviton mass is $m_g < 4.7\times 10^{-23}\,{\rm{eV}}$ with $30M_{\odot}$ - $30M_{\odot}$ BH binary at $z=0.05$. This is already close to the graviton mass bound from aLIGO, $m_g < 7.7\times 10^{-23}\,{\rm{eV}}$ \cite{GW170104:detection}, and we cannot expect the significant improvement of the graviton mass bound in aLIGO era. For the constant $\delta_g \mu$ model, since $\delta_g$ and $t_{\rm c}$ are completely degenerated, we need to impose a prior on $t_{\rm c}$, which would be obtained from the observation of an electromagnetic transient counterpart to a GW event. Once $t_{\rm c}$ information is obtained, $\delta_g$ can be determined at a precision of $\delta_g  \sim 1.2 \times 10^{-17}$, independent of masses of a GW source.


We already had a GW event with its source redshift from an electromagnetic transient counterpart and an identified host galaxy, GW170817/GRB170817A. This event enabled us to constrain the GW speed so tightly. In a couple of years, such events are expected to be detected more frequently by the GW detector network. Therefore, our universal framework for generalized GW propagation will be a useful tool to constrain gravity theories beyond GR.

\appendix
\section{PhenomD waveform}
\label{app:IMR}

The PhenomD waveform \cite{Khan:2016PRD} is composed of three parts (inspiral, intermediate, and merger-ringdown phases) and is given by
\begin{equation}
h_{\rm GR} = {\cal G}_I A_{\rm IMR}\, e^{i \phi_{\rm IMR}} \;, \nonumber \\
\end{equation}
with
\begin{equation}
A_{\rm IMR} =  \left\{
\begin{array}{ll|} 
\displaystyle
A_{\rm ins}  \quad \quad f \leq f_{a1}
\\ \\
\displaystyle 
A_{\rm int} \quad \quad f_{a1} < f \leq f_{a2}
\\ \\
\displaystyle 
A_{\rm MR} \quad \quad f_{a2} < f
\end{array}
\right. \;, 
\end{equation}
\begin{equation}
\phi_{\rm IMR} =  \left\{
\begin{array}{ll|} 
\displaystyle
\phi_{\rm ins,E} + \phi_{\rm ins,L}  \quad f \leq f_{p1}
\\ \\
\displaystyle 
\phi_{\rm int} \quad \quad f_{p1} < f \leq f_{p2}
\\ \\
\displaystyle 
\phi_{\rm MR} \quad \quad f_{p2} < f
\end{array}
\right. \;,
\end{equation}
and ${\cal G}_I$ the geometrical factor including detector response functions and the relative orientations of $I$th detector and a GW source. Each part is described by 
\begin{align}
A_0 &= \frac{1}{\sqrt{6} \pi^{2/3} d_{\rm L}} {\cal M}^{5/6} f^{-7/6} \;,
\label{eq:A0} \\
A_{\rm ins} &= A_0 \left\{ \sum_{i=0}^6 {\cal{A}}_i (\pi f)^{i/3} + \sum_{i=1}^{3} \rho_i f^{(i+6)/3} \right\} \;, \\
A_{\rm int} &= A_0 \sum_{i=0}^{4} \delta_i f^i \;, \\
A_{\rm MR} &= A_0 \gamma_1 \frac{\gamma_3 f_{\rm damp}}{(f-f_{\rm RD})^2 + \gamma_3^2 f_{\rm damp}^2} e^{-\frac{\gamma_2 (f-f_{\rm RD})}{\gamma_3 f_{\rm damp}}} \;, \\
\phi_{\rm ins,E} &= 2 \pi f t_{\rm c} -\phi_c -\pi/4 \nonumber \\
&+ \frac{3}{128} \left( \pi {\cal{M}} f \right)^{-5/3} \sum_{i=0}^7 \varphi_i (\pi M f)^{i/3} \;, \label{eq:PNphase} \\
\phi_{\rm ins,L} &= \frac{1}{\eta} \left\{ \sigma_0 + \sum_{i=1} \frac{3}{i+2}\sigma_i f^{(i+2)/3} \right\} \;, \\
\phi_{\rm int} &= \frac{1}{\eta} \left\{ \beta_0 + \beta_1 f + \beta_2 \log f - \frac{\beta_3}{3} f^{-3} \right\} \;, \\
\phi_{\rm MR} &= \frac{1}{\eta} \left\{ \alpha_0 + \alpha_1 f - \alpha_2 f^{-1}  + \frac{4}{3} \alpha_3 f^{3/4} \right. \nonumber \\
& \left. + \alpha_4 \tan^{-1} \left( \frac{f-\alpha_5 f_{\rm RD}}{f_{\rm damp}} \right) \right\} \;,
\end{align}
where $M$ is the total mass and is related to the chirp mass as $M={\cal M}\eta^{-3/5}$, $d_{\rm L}$ is luminosity distance. The explicit expressions of the coefficients $\varphi_i, \alpha_i, \beta_i, \gamma_i, \delta_i, \sigma_i, \rho_i, {\cal A}_i$ are given in \cite{Khan:2016PRD} and some of them are fixed from matching conditions between different parts of the waveform. The transition frequencies of the waveform are $f_{a1}=0.014 f_M$ and $f_{a2}=f_{\rm peak}$ for amplitude and $f_{p1}=0.018 f_M$ and $f_{p2}=0.5 f_{\rm RD}$ for phase, where
\begin{align}
f_M &= M^{-1} \approx 440\,{\rm Hz} \left( \frac{10\,M_{\odot}}{M} \right) \;, \\
f_{\rm peak} &= \left| f_{\rm RD} + \frac{f_{\rm damp} \gamma_3 (\sqrt{1-\gamma_2^2}-1) }{\gamma_2} \right| \;, \\
f_{\rm RD} &= \frac{f_M}{2\pi} \left\{ 1.5251-1.1568 (1-a_f^{\rm eff})^{0.1292} \right\} \;, \\
f_{\rm damp} &= \frac{f_{\rm RD}}{2Q} \;, \\
Q &= 0.7000+ 1.4187 (1-a_f^{\rm eff})^{-0.4990} \;, \\
a_f^{\rm eff} &= S + 2\sqrt{3} \eta -4.399 \eta^2 +9.397 \eta^3 -13.181 \eta^4 \nonumber \\
&+(-0.085 S+0.101 S^2 -1.355 S^3 -0.868 S^4) \eta \nonumber \\
&+ (-5.837 S-2.097 S^2 +4.109 S^3 +2.064 S^4) \eta^2 \;, \\
S &\equiv \frac{S_1+S_2}{M^2} \nonumber \\
&= (1-2\eta) \chi_{\rm s} + \sqrt{1-4\eta}\, \chi_{\rm a} \;.
\end{align}
Here $a_f^{\rm eff}$, $S$ are from \cite{Husa:2015PRD} and $f_{\rm damp}$, $f_{\rm RD}$, and $Q$ are from \cite{Santamaria:2010PRD}.

\begin{acknowledgments}
A.N. was supported by NSF CAREER Grant No. PHY-1055103, the H2020-MSCA-RISE-2015 Grant No. StronGrHEP-690904, and JSPS KAKENHI Grant Number JP17H06358.
\end{acknowledgments}

\bibliography{/Volumes/USB-MEMORY/my-research/bibliography}

\end{document}